%% file: main.tex
\documentclass[%
reprint,
superscriptaddress,
nofootinbib,
amsmath,amssymb,
aps,
prl,
tightenlines,
nolongbibliography,
eprint,
floatfix,
]{revtex4-2}

\input{preamble.tex}
\begin{document}

\preprint{APS/123-QED}

\title{Breakup dynamics in primary jet atomization using mesh- and interface- refined Cahn-Hilliard Navier-Stokes}

\author{Makrand A. Khanwale}
\thanks{Corresponding authors}
\affiliation{Center for Turbulence Research, Stanford University, Stanford, CA 94305, USA}%
\author{Kumar Saurabh}%
\affiliation{Department of Mechanical Engineering, Iowa State University, Ames, IA 50010, USA}%
\author{Masado Ishii}
\affiliation{School of Computing, The University of Utah, Salt Lake City, UT 84112, USA}%
\author{Hari Sundar}
\affiliation{School of Computing, The University of Utah, Salt Lake City, UT 84112, USA}%
\author{Baskar Ganapathysubramanian}
\thanks{Corresponding authors}
\affiliation{Department of Mechanical Engineering, Iowa State University, Ames, IA 50010, USA}%




\begin{abstract}

We present a technique to perform interface-resolved simulations of complex breakup dynamics in two-phase flows using the Cahn-Hilliard Navier-Stokes equations. The method dynamically decreases the interface thickness parameter in relevant regions and simultaneously increases local mesh resolution, preventing numerical artifacts. We perform a detailed numerical simulation of pulsed jet atomization that shows a complex cascade of break-up mechanisms involving sheet rupture and filament formation. To understand the effect of refinement on the breakup, we analyze the droplet size distribution. The proposed approach opens up resolved simulations for various multiphase flow phenomena.
\end{abstract}

\maketitle


\section{Introduction}
A fluid jet injected into a lighter-density fluid (like air) exhibits a rich tapestry of flow physics, including the rupture of fluid films into filaments, the breakup of filaments into droplets, and a cascade of droplet breakup and coalescence. Understanding this jet atomization process can transform how we interact with, design, and control a wide array of natural and engineered systems, for example, combustion, printing, coating, and spraying  operations. 

\begin{figure}[!ht]
	\begin{center}
	\begin{tabular}{c} 
		\subfigure [] {
            \centering
			{\includegraphics[width=0.5\linewidth, 
            trim=6.5in 0 4.0in 0,clip]{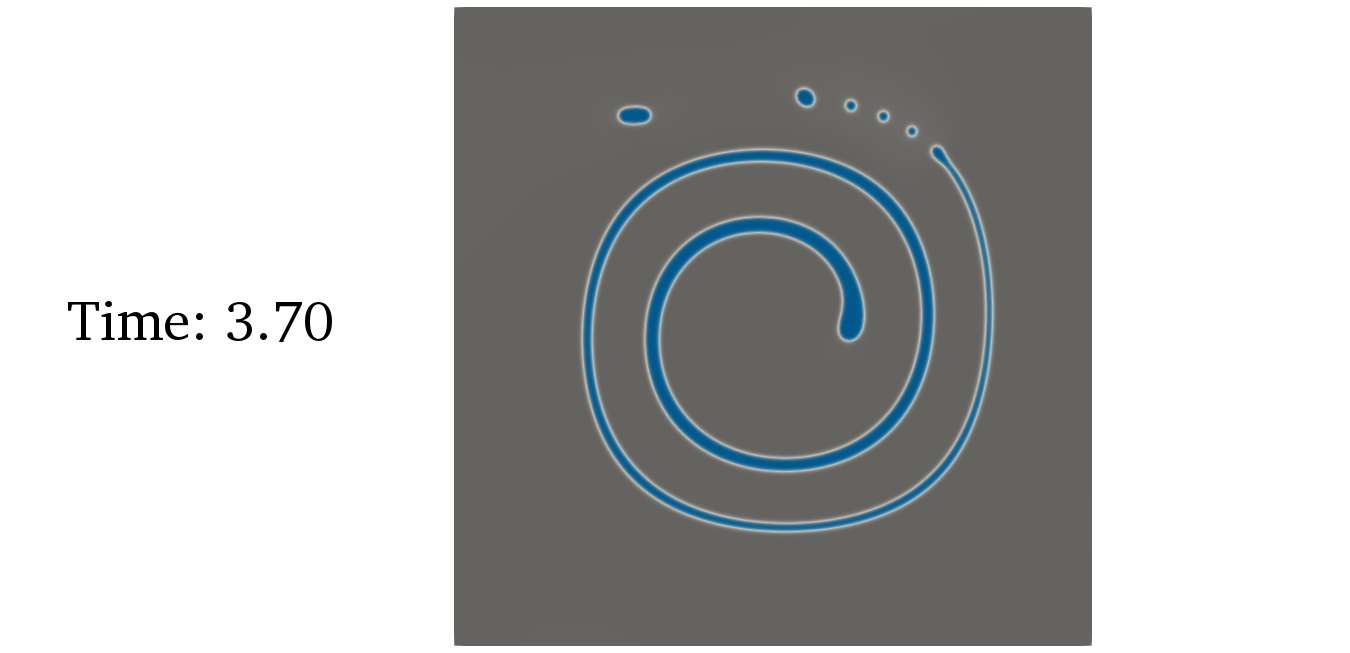}}
			\label{subfig:rt_snap_1}
		} \\ 
		\subfigure [] {
            \centering
            {\includegraphics[width=0.5\linewidth, 
            trim=6.5in 0 4.0in 0,clip]{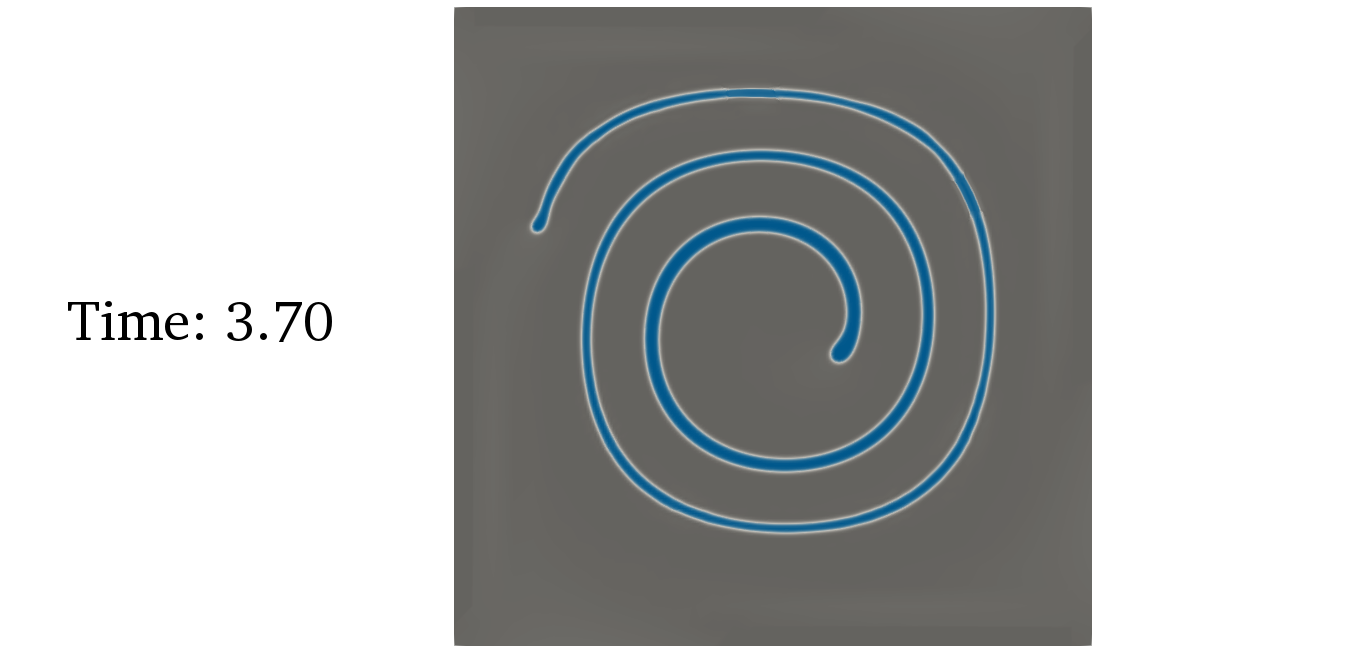}}
			\label{subfig:rt_snap_2}
		}  
	\end{tabular}
	\caption{Consider a droplet in swirling flow. The swirling flow gradually stretches out the droplet into a spiraling and thinning filament. Fig (a) is a simulation with constant interface thickness, $\epsilon$, and shows spurious breakup into droplets when the filament thickness becomes comparable to  $\epsilon$. In contrast, Fig (b) is a simulation where $\epsilon$ is always smaller than the local filament thickness. See animations for the evolution in both cases in the supplementary video}
	\label{fig:swirling_flow}
	\end{center}
	\vspace{-0.35in}
\end{figure}

Jet atomization -- the dynamics of generation of micro-droplets from a compact fluid core via the breakup of unstable liquid sheets and filaments~\citep{Villermaux2007, Majumdar2018} -- has proved notoriously challenging to understand, primarily due to the presence of a wide range of spatial scales~\citep{Gorokhovski2008}. For instance, if we consider the liquid jet diameter $D_i$, atomization produces fluid sheets that break up into filaments. These thin sheets and long filaments have one (sheets) or two (filaments) of their characteristic dimensions $100 \times$ smaller than $D_i$. The subsequent breakup of the filaments produces droplets as much as $1000 \times$ smaller than $D_i$. The ensuing 3D structures span nine orders of magnitude in volume and must be resolved accurately to capture the droplet distribution dynamics. Computationally resolving the spatio-temporal dynamics of these multi-scale structures calls for approaches that accurately track the fluid-air interface. 

\begin{figure*}
	\centering
	\includegraphics[width=\linewidth]{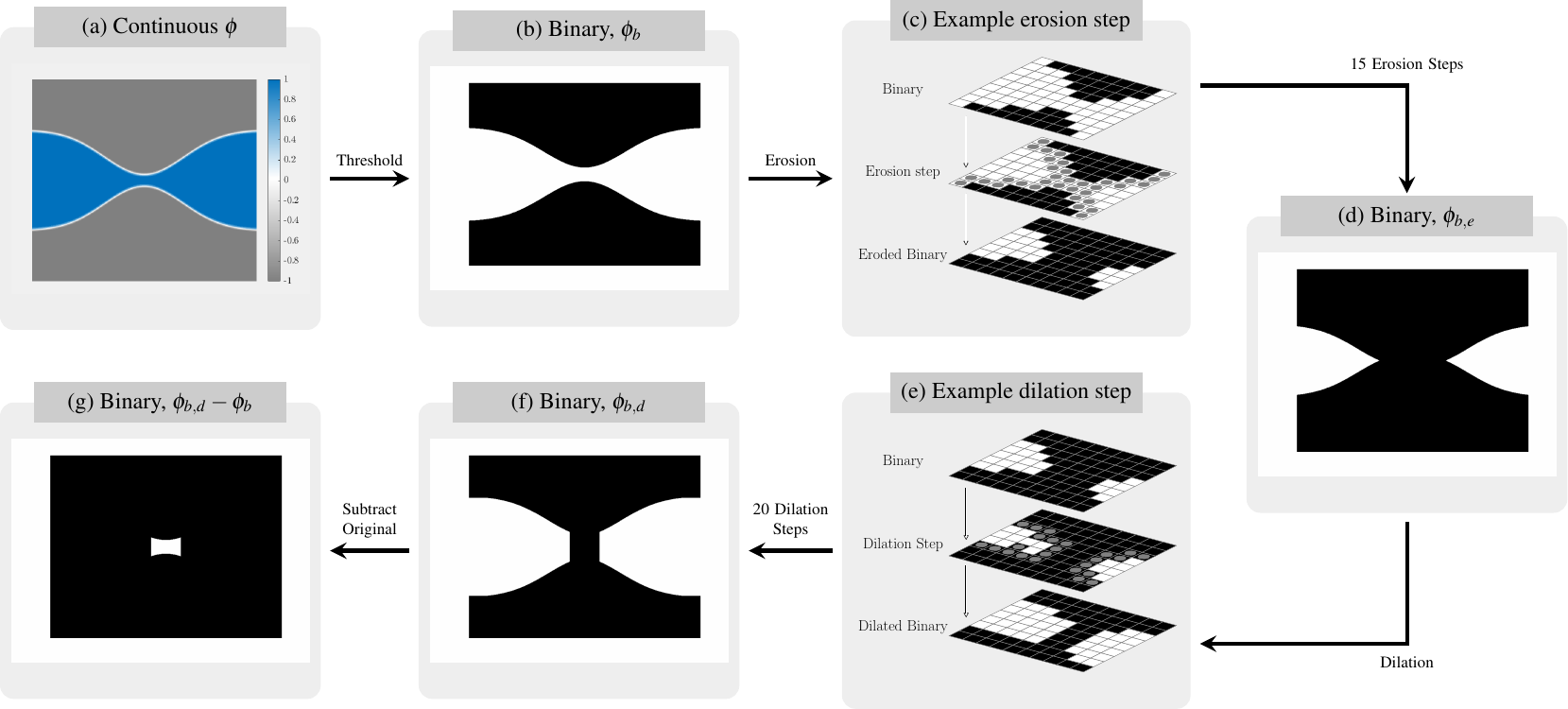}
	\caption{Flow chart describing the algorithm for detecting thin/small interface features. (a) Start with the continuous field $\phi$; (b) Threshold to produce a binary image; (c) An illustration of one step of erosion, the top image shows a sample binary with white cells equal to 1, black cells equal to the value of 0. Gray dots in the middle picture shows cells identified (with at least one neighbor of 0) to be converted to 0. The bottom picture shows binary after cells with Gray dots turned to 0; (d) Binary after 15 steps of erosion on $\phi_{b}$; (e) An illustration of one step of dilation, the top image shows a sample eroded binary. Gray dots in the middle picture shows cells identified (with at least one neighbor 1) to be converted to 1. The bottom picture shows binary after cells with Gray dots turned to 1; (f) Binary after 20 steps of dilation on $\phi_{b,d}$;(g) Binary after original $\phi_{b}$ is subtracted from $\phi_{b,d}$ which shows the identified small filament region.}
	\label{fig:schematic}
	\vspace{-0.2in}
\end{figure*}

In diffuse interface simulations of multi-phase flow, the atomically thin fluid-air interface is replaced with an interface having a much larger and constant thickness, $\epsilon$, for computational efficacy. As long as the relevant length scales are larger than $\epsilon$, such approaches work~\citep{Yue2007}. However, in phenomena like jet atomization, thin filaments and droplets have characteristic dimensions that become comparable to $\epsilon$; thus, the standard approach produces spurious artifacts like mass loss and nonphysical breakup (see \cref{subfig:rt_snap_1}, and supplementary video of \cref{subfig:rt_snap_1}). While one could choose a very small $\epsilon$, the associated computational cost makes this infeasible. In this paper, our first  modeling 
advance is to replace a \textit{constant} interface thickness model with a \textit{locally adaptive interface thickness}. This ensures that the (local) interface thickness, $\epsilon$, is always smaller than the \textit{local length scales of interest} (see \cref{subfig:rt_snap_2}), thus allowing interface tracking approaches to reliably capture multi-scale features in a computationally efficient manner.
Note that automatically identifying the local regions of interest where $\epsilon$ has to be adapted is itself non-trivial. We leverage concepts from image processing (see~\cref{fig:schematic}) to build an efficient approach to do so, which constitutes our second modeling advance. While the first modeling advance is specific to diffuse interface approaches, the second modeling advance applies to all interface capturing methods, including sharp interface approaches. We briefly discuss the current challenges of both sharp and diffuse interface approaches next and how the current work (as illustrated in \cref{fig:swirling_flow}) resolves these challenges.

Sharp interface approaches have been very successful for high-fidelity simulations of turbulent multiphase flows, including atomization~\citep{Fuster2009, Tomar2010, Herrmann2010, Shinjo2011, Hasslberger2019, Pairetti2020}.
However -- as in the case of the diffuse interface approaches -- when the interfacial features (droplets/filaments) are comparable to the grid size, the numerical breakup is observed as the method tries to conserve volume; this effect is also known as numerical surface tension~\citep{Gorokhovski2008}. Active research continues towards methods -- for example, interface reconstruction~\citep{Lu2018, Chiodi2020}, moment-of-fluid~\citep{Jemison2015}, manifold death~\citep{Chirco2022} -- that resolve this issue. These methods, however, remain computationally complex because the interface quality is determined purely by the mesh resolution. 

Diffuse interface approaches consider a computationally smeared out fluid-fluid interface with thickness $\epsilon$. By keeping $\epsilon$ relatively large, comparable results can be produced under coarser mesh resolutions. These methods also avoid any interface reconstruction process, provide a thermodynamically consistent way to handle surface tension~\citep{Anderson1998}, and are endowed with desirable numerical properties like smoothness and energy stability.
However, when interfacial features have a comparable length scale to $\epsilon$, diffuse interface methods also suffer from artificial breakup and mass loss. We leverage that diffuse interface approaches allow us to disentangle the interface thickness, $\epsilon$, from the mesh resolution. When the interfacial feature length scales, $r$ become comparable to the numerical interface thickness ($\epsilon/r \sim \mathcal{O}(1)$), we locally reduce $\epsilon$. Subsequently, a more refined mesh resolution is needed only locally to accurately capture this interface with lower $\epsilon$. This is especially appealing over current approaches of decreasing $\epsilon/r$ everywhere, which are prohibitively expensive even with adaptive mesh refinement. 

Note that the ability to efficiently identify regions of interest for subsequent local refinement is readily applicable to both sharp and diffuse interface approaches, potentially improving a large array of interface-capturing methods. This technique -- implemented into our massively parallel, adaptive mesh refined, multi-phase flow framework and described in detail in a companion methods paper~\citep{Kumar2022ar} (see also SI)-- allows us to perform one of the most detailed simulations of primary jet atomization at $64\times$ the previous best possible resolution~\citep{Pairetti2020}. Specifically, we use a thermodynamically consistent diffuse interface method that uses Cahn-Hilliard Navier-Stokes (CHNS) as detailed in~\citep{Khanwale2021}.






\section{Identification of regions of interest}
\label{sec:ident}

We describe the algorithm (see~\cref{fig:schematic}) for identifying regions such as filaments and small droplets where $\epsilon/r$ can become $\sim \mathcal{O}(1)$. Consider a two-fluid system. The two fluid phases are described by a phase field variable $\phi$, which varies continuously in $[-1, 1]$, with -1 and 1 corresponding to the pure phases; see panel (a) of~\cref{fig:schematic}. The central thin part in the figure is a region of interest. We perform a sequence of efficient image processing steps to identify this region. We first threshold the continuous field, $\phi$ into a binary field, $\phi_b$ \footnote{$\phi \geq 0.8 = 1$ and $\phi \le 0.8 = 0$ with each cell being either 1 or 0} (see panel (b) in~\cref{fig:schematic}). We then perform a predetermined number of morphological \textit{erosion} steps (see panel (c)~\cref{fig:schematic}) on the binary image~\footnote{One erosion step converts cells 
with a value of 1 to 0 if at least one of the neighbors of the cell its neighbor is 0} to get $\phi_{b,e}$ (see panel (d)~\cref{fig:schematic}). 

We could subtract $\phi_{b,e}$ from the original binary $\phi_b$ to find the filament region, but the erosion steps also remove parts of the bigger structure. To recover the part of the bigger structure without recovering the small feature, we perform a predetermined number of morphological \textit{dilation} steps \footnote{The dilatation step involves converting the cell value from 0 to 1 if at least one of its neighbors is 1.} on the eroded binary $\phi_{b,e}$ to get the dilated binary $\phi_{b,d}$ (panel (e) in~\cref{fig:schematic}). We perform slightly more dilation steps than erosion steps to completely recover the larger structure (see panel (f)~\cref{fig:schematic}). Finally, we subtract $\phi_{b}$ from $\phi_{b,d}$ to get a binary field with non-zero values in the regions representing the small features of interest (see panel (g) of \cref{fig:schematic}). 

We choose such a computational graphics-based approach for several reasons: (a) \textit{computational complexity}: all steps detailed above can be performed using algorithms exhibiting $\mathcal{O}(N)$ complexity, where $N$ is the total number of cells in the mesh, (b) \textit{parallel scalability}: these algorithms can be implemented to scale efficiently on a large number of processors. We refer readers to the companion method papers~\citep{Kumar2022ar} for a detailed discussion of parallel deployment on octree meshes, (c) \textit{algorithmic alignment:} these algorithms (re)use the data structures and communication constructs present in our existing, optimized codebase~\citep{saurabh2021scalable, Khanwale2021}. We note that other approaches, for instance, connected component analysis, can also be applied. But these come at a higher computational cost, specifically in a distributed setting, and our experiments indicated that they could not identify all features of interest (like long thin filaments attached to larger structures). 

To demonstrate the capability of the identification algorithm, we choose a popular problem of a drop in swirling flow, which is often used to test interface methods, especially under resolution constraints~\citep{Bell1989, Rider1998, Tryggvason2011}. A circle with radius 0.15, centered at $(0.5, 0.75)$ is placed inside a unit box (origin at bottom left corner) and undergoes advection in a swirling velocity field given by the stream function $\varphi(x,y,t) = \frac{1}{\pi}\sin^2(\pi x) \sin^2(\pi y)$. We set $\epsilon$ to 0.0025. The circle advects and deforms to form concentric thin filaments. With increasing time, the thickness of the filament becomes comparable to the interface thickness, and we see breakup and coarsening for CHNS, \cref{subfig:rt_snap_1}. In sharp interface methods like VOF, as the filament grows thinner, they break into droplets under limited mesh resolution, thus showing a "numerical" surface tension~(see~Fig. 5.21 in~\citep{Tryggvason2011}). Therefore, we see a breakup for both sharp and diffuse interface methods due to limiting mesh resolution.  Our approach detects thin filaments' regions as they develop. We adaptively reduce the interface thickness $\epsilon$ in those regions (to 0.001) and locally use a higher mesh resolution. Consequently, the numerical breakup is prevented.  \Cref{fig:swirling_flow} shows the comparison between the case of constant $\epsilon$ (panel (a)) compared to the case where we dynamically decrease $\epsilon$ (panel (b)). We provide videos of simulations for significantly longer times (up to $t=8$) in the supplementary files, which illustrates the utility of this approach. 

\begin{figure*}
	\centering
         \includegraphics[width=0.9\linewidth]{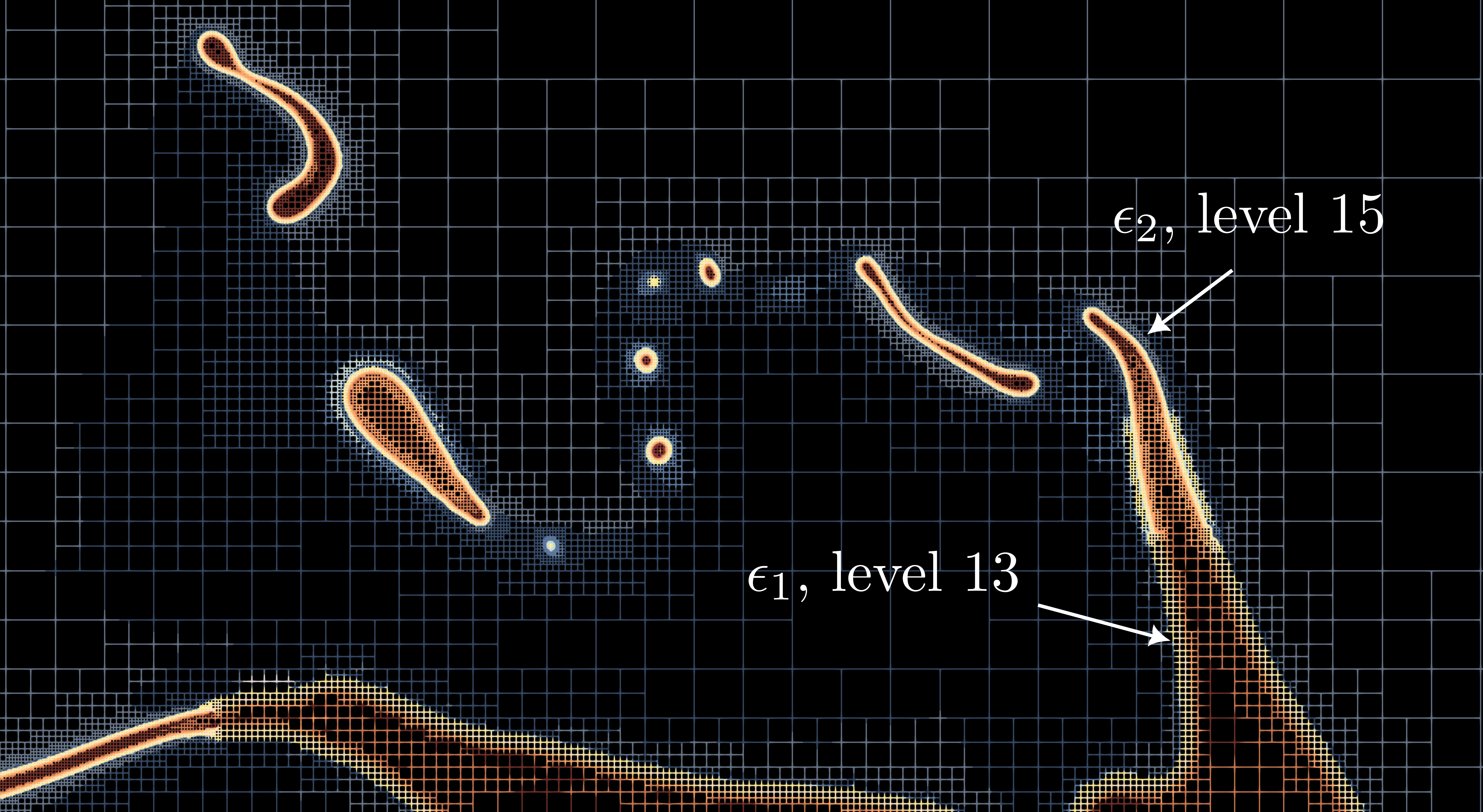} 
	\caption{ A snapshot of the mesh overlay on the zoomed-in 2D plane view of the diffuse interface at the mid-z plane.    
    }
    \label{fig:mesh}
\end{figure*}
\section{Primary jet atomization}
We deploy this approach on the challenging problem of pulsed primary jet atomization  from~\citet{Pairetti2020} with high Reynolds numbers ($Re_{gas} = 25000$, $Re_{liquid} = 5800$)~\footnote{$Re_{liquid} = \rho_{liquid}U_{i}D_{i}/\mu_{liquid}$, $Re_{gas} = \rho_{gas}U_{i}D_{i}/\mu_{gas}$, $We_{gas} = \rho_{gas}U^2_{i}D_{i}/\sigma = 430$, and $We_{gas} = \rho_{liquid}U^2_{i}D_{i}/\sigma = 11600$}. A liquid jet with the properties of diesel is pulsed into a large box filled with air. The max inlet velocity, $U_{i}$ is 100 \si{m/s} with the diameter of the injection $D_{i} = \SI{100}{\micro\metre}$. All simulation parameters and boundary conditions are identical to those in~~\citep{Pairetti2020} and are detailed in the supplementary material. 
Analysis by \citet{Pairetti2020} revealed that mesh resolution significantly affects breakup statistics. They showed that even using the finest mesh refinement level of 13 -- corresponding to the cell size of $30D_{i}/2^{13}$= \SI{0.36}{\micro\metre} -- still produces structures of the same order as the mesh. 

Here, our approach allows us to reach a mesh refinement of 15, corresponding to a finest cell size of $30D_{i}/2^{15}$ = \SI{0.0915}{\micro\metre}.  We believe this to be one of the finest resolution simulations of primary jet atomization. Our simulation spanned 120000 timesteps reaching 6.3 $\si{\micro\second}$, and required over $200,000$ node hours on the supercomputer,  \href{https://www.tacc.utexas.edu/systems/frontera}{TACC Frontera}~\footnote{We estimate that without dynamic $\epsilon$, using level 15 everywhere on the interface would result in a $20 \times$ more expensive simulation\citep{Kumar2022ar}}.

The dynamics are characterized by two length scales, Kolmogorov and Hinze. The Kolmogorov length scales represent the size of the smallest turbulent structures (in both liquid $\eta_l$ and air $\eta_g$), and the Hinze scale ($\xi$) represent the size of the biggest drop that does not suffer breakup by turbulent fluctuations~\footnote{$\eta_l = D_{i}/Re_{l}^{3/4}$, $\eta_g = D_{i}/Re_{g}^{3/4}$, $\epsilon = \nu_{gas}^3/\eta_{gas}^4$, $\xi = 0.75 \left(\sigma/\rho_{gas}\right)^{3/5}\left(\epsilon/\rho_{gas}\right)^{-2/5}$}. Here, $\eta_l = \SI{0.131}{\micro\metre}$, $\eta_g = \SI{0.078}{\micro\metre}$, and $\xi = \SI{9.71}{\micro\metre}$. To achieve DNS resolution, the mesh size $h \leq 2\min (\eta_l, \eta_g)$. We achieve DNS resolution near the dynamically refined interface. The fine resolution near the interface resolves the fine-scale shear instabilities near the interface, which impact breakup mechanisms. 

We start with a non-dimensional interface thickness $\epsilon_1 = 0.001$. Using the algorithm in~\cref{fig:schematic}, we locally decrease the interface thickness to $\epsilon_2 = 0.00075$. The refinement in these regions changes from level 13 to level 15. \Cref{fig:mesh} shows the mesh overlaid on the diffuse interface on a zoomed-in view of a mid-z slice near the top of the jet. Notice that our algorithm detects small droplets and sheets and imposes a smaller $\epsilon$ in these thin regions, with a finer resolution (level 15) only in these regions. Such a selected decrease in $\epsilon$ allows us to prevent numerical breakup and Ostwald ripening. As seen in \Cref{fig:mesh}, we impose an extended region of refinement near the interface~\footnote{either level 13 if $\epsilon$ = 0.001 or level 15 if the $\epsilon$ = 0.00075.}. Thus, any droplet formed with a length scale less than the Hinze length does not break up further through turbulent fluctuations. 

\begin{figure*}
    \centering
    \includegraphics[width=0.75\linewidth]{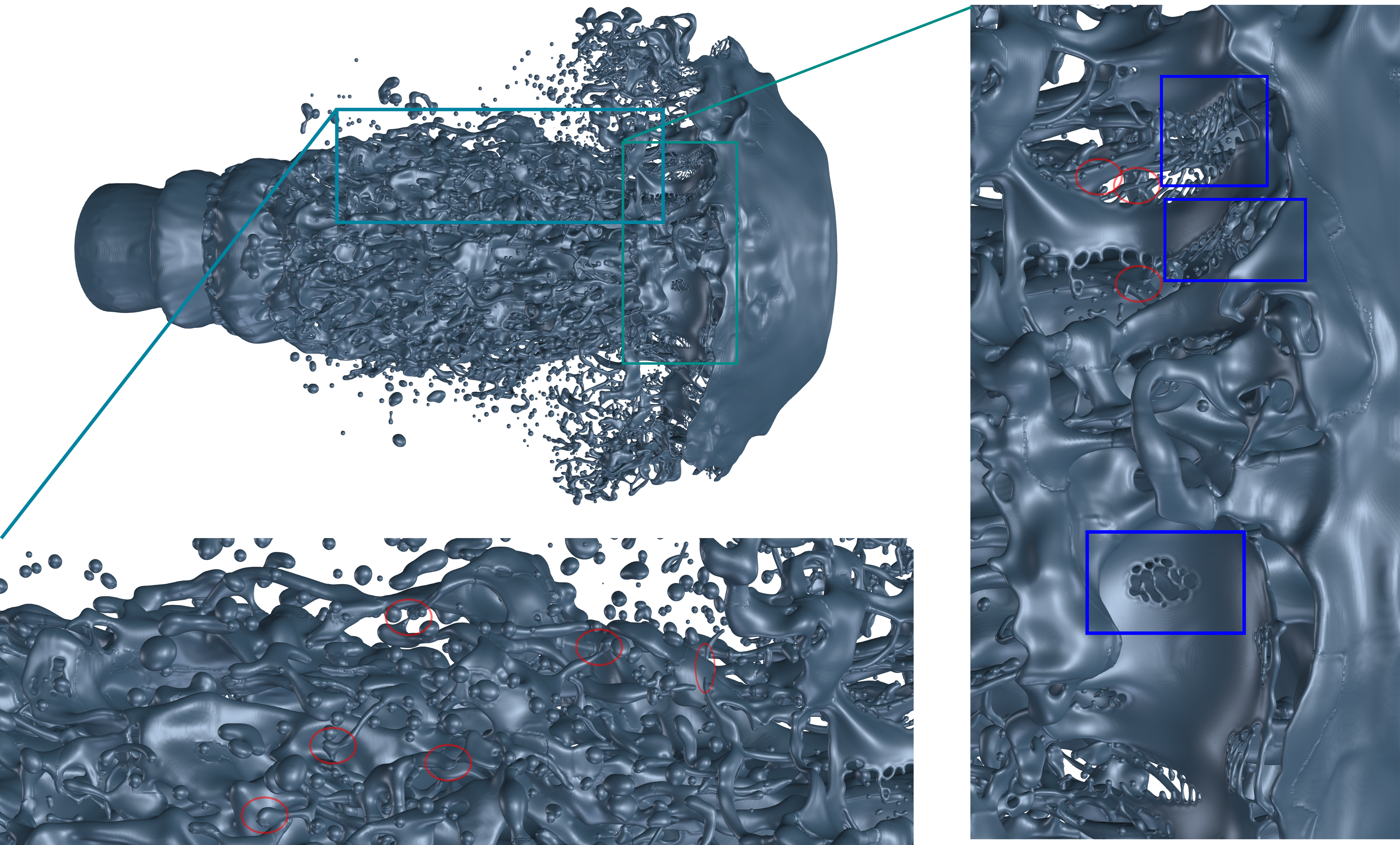}
    \caption{Isocontour of diffuse interface parameter showing the liquid-air interface at $\SI{6.2}{\micro\second}$. We present two zoomed-in views of the jet, one focusing on sheet rupture on the right and the second focusing on filament breakup below. Examples of sheet ruptures are marked with blue rectangles, and examples of filament breakup are marked with red ellipses.  Please see supplementary information for a full animation of the evolution of the jet. }
    \label{fig:jet_end}
\end{figure*}

This strategy allows tracking both droplets as well as thin sheets and filaments. We discovered that (see \Cref{fig:jet_end} and accompanying video) sheets are formed as early as \SI{1.68}{\micro\second}. This is in contrast to lower resolution studies that report  sheet formation around \SI{5}{\micro\second}~\citep{Pairetti2020}. As the jet evolves, we notice a cascade of sheet ruptures near the jet's tip, forming filaments~(see~\cref{fig:jet_end}). This destabilization and rupture is typically via the Kelvin-Helmholtz instability. This rupture can be seen in the blue bounding boxes in \Cref{fig:jet_end}(right). To our best knowledge, this is the first time that such rupture dynamics have been computationally captured. The filaments then fragment to form droplets through the Rayleigh-Plateau instability. Representative filament fragmentation events are indicated with red bounding boxes in \Cref{fig:jet_end}(bottom). Capturing this cascade of breakup phenomena is necessary for correctly computing the breakup statistics.

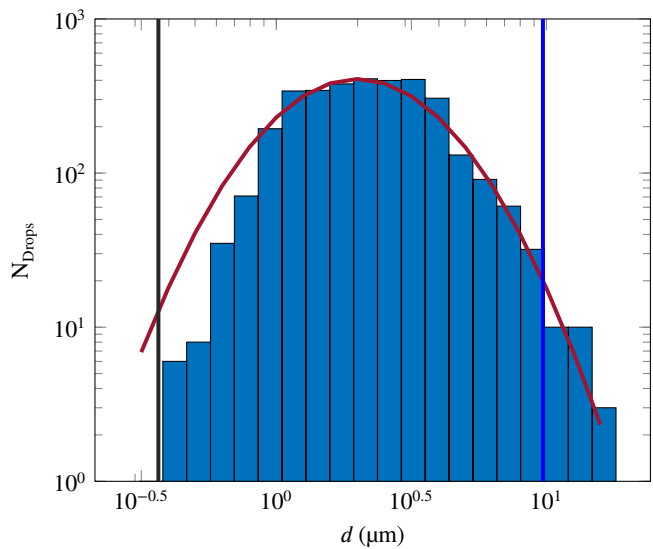
\begin{figure}
	\centering
		\begin{tikzpicture}
			\centering
			\begin{axis}[%
				width=0.5\textwidth,
				bar shift auto,
				log origin=infty,
				xlabel={$d\left(\SI{}{\micro\metre}\right)$},
				ymode=log,
                xmode=log,
                extra x ticks = {10^-0.5, 10^0.5},
                ymin=1,
				ymax=1000,
				yminorticks=true,
                xminorticks=true,
                ylabel={$\text{N}_{\text{Drops}}$},
				axis background/.style={fill=white}
				]
				\addplot[ybar, 
                bar width =1.224616199,
				fill=mycolor1, 
				draw=black, 
				area legend] 
				table[row sep=crcr,
                    x index=0,
                    x expr={10^(\thisrowno{0})}
                    ] {%
					-0.464549368592872	1\\
					-0.376216418760339	6\\
					-0.287883468927806	8\\
					-0.199550519095273	35\\
					-0.11121756926274	71\\
					-0.0228846194302074	194\\
					0.0654483304023254	341\\
					0.153781280234858	344\\
					0.242114230067391	379\\
					0.330447179899924	409\\
					0.418780129732457	399\\
					0.50711307956499	405\\
					0.595446029397523	306\\
					0.683778979230055	131\\
					0.772111929062588	91\\
					0.860444878895121	61\\
					0.948777828727654	32\\
					1.03711077856019	10\\
					1.12544372839272	10\\
					1.21377667822525	3\\
				};
				\addplot [
				color=mycolor2, 
				line width=1.5pt, 
				forget plot]
				table[row sep=crcr,
                        x index=0,
                        x expr={10^(\thisrowno{0})}
                    ]{%
					-0.5	6.88561671342292\\
					-0.4	17.9226872546738\\
					-0.3	41.0647047508234\\
					-0.2	82.8208299841944\\
					-0.1	147.033314660718\\
					0	229.772057435491\\
					0.1	316.070507562495\\
					0.2	382.715481322818\\
					0.3	407.918481710465\\
					0.4	382.715481322818\\
					0.5	316.070507562495\\
					0.6	229.772057435491\\
					0.7	147.033314660718\\
					0.8	82.8208299841944\\
					0.9	41.0647047508235\\
					1	17.9226872546738\\
					1.1	6.88561671342294\\
					1.2	2.32856227596307\\
				};
				\addplot [color=blue, line width=1.5pt, forget plot]
				table[row sep=crcr,
                        x index=0,
                        x expr={10^(\thisrowno{0})}
                    ]{%
					0.98721923	1\\
					0.98721923	1000\\
				};
				\addplot [color=white!15!black, line width=1.5pt, forget plot]
				table[row sep=crcr,
                        x index=0,
                        x expr={10^(\thisrowno{0})}
                    ]{%
					-0.4365	1\\
					-0.4365	1000\\
				};
			\end{axis}
		\end{tikzpicture}%
\caption{ 
  \textit{Droplet statistics in jet atomization at $\SI{6.21}{\micro\second}$: } A histogram of the number of droplets for a range of logarithms of the length scales (in \SI{}{\micro\metre}) representing the droplets. We fit the Log-normal PDF with a mean of $10^{0.3}$ and a standard variation of $10^{0.28}$ shown with the red curve. The black vertical line shows the thinnest interface thickness used $\epsilon_2 = 4h$, where $h$ is the size of the smallest mesh element. The blue vertical line represents the Hinze scale $\xi$, and the black vertical line represents the diffuse interface thickness $\epsilon_2$.
}
\label{fig:droppdf}
\end{figure}

We construct breakup statistics by computing the ensuing structures' representative length scale ($d$). We compute the volume ($V_d$) of each connected components, and then calculate $d = \left(6V_d/\pi\right)^{1/3}$. Breakup statistics are constructed at the final time of our simulation, $\SI{6.3}{\micro\second}$. \Cref{fig:droppdf} shows this size distribution of droplets, along with a best-fit log-normal distribution. We point out several observations: 
\begin{itemize}
    \item The mean value of the histogram is much larger than the interface thickness (indicated by the black vertical line). Additionally, most of the histogram is smaller than the Hinze scale $\xi$; therefore, the bulk of the distribution consists of droplets that will not further break up due to turbulence. The log-normal fit to droplet distribution has also been observed in other jet atomization simulations~(see \cite{Herrmann2010, Pairetti2020}). The distribution suggests that all the features in the distribution are well resolved. 
    \item The histogram is to the right of the interface thickness. This indicates that all the resolved drops are larger than the interface thickness. Furthermore, drops smaller than the interface thickness are not resolved by the method and do not pollute the statistics. In the accompanying movie, we observe that droplets with sizes around the order of the interface thickness get absorbed into nearby larger structures due to the coarsening nature of the Cahn-Hilliard equation, which is why we do not see tails in the droplet size distribution near the interface thickness. One could resolve these smaller droplets by deploying a multi-level interface thickness refinement. That is, the interface thickness can be recursively refined to capture smaller and smaller droplets.
    \item  We notice no pile-up of droplets with a length scale smaller than the interface thickness. This indicates that no artificial droplets result from numerical breakup due to low mesh resolution, unlike other state-of-the-art volume-of-fluid simulations. 
\end{itemize}

The source code and associated datasets of the simulation are publicly available. We anticipate this to produce lively discussions and analysis of the data. We continue to analyze the hydrodynamical conditions of sheet rupture and plan to deploy another level of interface refinement to capture smaller droplets. A follow-up paper will present an additional analysis of the hydrodynamic conditions of different breakup events and the turbulent statistics of the flow. While the computational approach has been used to model an important canonical multi-flow physics problem, this approach has wide applicability to other fluid mechanics phenomena, like low Reynolds number ligament breakup.

\begin{acknowledgments}
    \textit{Acknowledgments}: We acknowledge fruitful conversations with Prof. Ali Mani and Prof. Marcus Herrmann. We thank Greg Foss and Dave Semeraro for the detailed and insightful visualizations. We thank TACC Frontera and XSEDE program for computational resources. 
\end{acknowledgments}

\bibliography{references}

\end{document}


\preprint{APS/123-QED}

\title{Supplemental Information
: Breakup dynamics in primary jet atomization using mesh- and interface- refined Cahn-Hilliard Navier-Stokes}

\author{Makrand A. Khanwale}
\thanks{Corresponding authors}
\affiliation{Center for Turbulence Research, Stanford University, Stanford, CA 94305, USA}%
\author{Kumar Saurabh}%
\affiliation{Department of Mechanical Engineering, Iowa State University, Ames, IA 50010, USA}%
\author{Masado Ishii}
\affiliation{School of Computing, The University of Utah, Salt Lake City, UT 84112, USA}%
\author{Hari Sundar}
\affiliation{School of Computing, The University of Utah, Salt Lake City, UT 84112, USA}%
\author{Baskar Ganapathysubramanian}
\thanks{Corresponding authors}
\affiliation{Department of Mechanical Engineering, Iowa State University, Ames, IA 50010, USA}%



\date{\today}

\begin{abstract}

\end{abstract}

\maketitle

\section{Governing Equations}
Consider a domain $\Omega$ with two immiscible fluids with $\rho_{+}$ ($\eta_{+}$ ) and $\rho_{-}$ ($\eta_{-}$) as the specific density (viscosity). 
Let $\phi$ be the smooth phase field that varies between $+1$ to $-1$ with $+1$ or $-1$ corresponding to either pure phase. Then the non-dimensional mixture density $\rho(\phi)$ and the non-dimensional mixture viscosity $\eta(\phi)$ defined as, 
\begin{align}
\rho(\phi) = \alpha\phi + \beta, \quad
\text{where} \quad \alpha = \frac{\rho_+ - \;\rho_-}{2\rho_+} 
\quad \text{and} \quad \beta = \frac{\rho_+ + \;\rho_-}{2\rho_+},\\
\eta(\phi) = \gamma\phi + \xi, \quad \text{where} \quad 
\gamma = \frac{\eta_+ - \;\eta_-}{2\eta_+} \quad \text{and}
\quad \xi = \frac{\eta_+ + \;\eta_-}{2\eta_+}.
\end{align}
Note that our non-dimensional form uses the specific density/viscosity of fluid 1 (+) as the non-dimensionalizing quantity. 

Let $\vec{v}$ be the volume averaged mixture velocity, $p$ the volume averaged pressure, and $\mu$ be the chemical potential.  Then the thermodynamically consistent Cahn-Hilliard Navier-Stokes equations~\citep{Abels2012} in their non-dimensional form are as follows:
\begin{align}
\begin{split}
	\text{Momentum Eqns:} & \quad \pd{\left(\rho(\phi) v_i\right)}{t} + \pd{\left(\rho(\phi)v_iv_j\right)}{x_j} + \frac{1}{Pe}\pd{\left(J_jv_i\right)}{x_j} +\frac{Cn}{We} \pd{}{x_j}\left({\pd{\phi}{x_i}\pd{\phi}{x_j}}\right) \\
	& \quad \quad \quad + 
	\frac{1}{We}\pd{p}{x_i} - \frac{1}{Re}\pd{}{x_j}\left({\eta(\phi)\pd{v_i}{x_j}}\right) - \frac{\rho(\phi)\hat{{g_i}}}{Fr} = 0, \label{eqn:nav_stokes} 
	\end{split} \\
	\text{Solenoidality:} & \quad \pd{v_i}{x_i} = 0, \label{eqn:solenoidality}\\
	\text{Continuity:} & \quad \pd{\rho(\phi)}{t} + \pd{\left(\rho(\phi)v_i\right)}{x_i}+
	\frac{1}{Pe} \pd{J_i}{x_i} = 0, \label{eqn:cont}\\
    \text{Cahn-Hilliard Eqn:} & \quad \pd{\phi}{t} + \pd{\left(v_i \phi\right)}{x_i} - \frac{1}{PeCn} \pd{}{x_i}\left(m(\phi){\pd{\mu}{x_i}}\right) = 0, \label{eqn:phi_eqn}
    \\
	\text{Chemical Potential:} & \quad \mu = \psi'(\phi) - Cn^2 \pd{}{x_i}\left({\pd{\phi}{x_i}}\right).\label{eqn:mu_eqn} 
\end{align}
$\hat{\vec{g}}$ is a unit vector defined as $\left(0, -1, 0\right)$ denoting the direction of gravity.  We use the polynomial form of the free energy density $\psi(\phi(\vec{x}))$ defined as follows:
\begin{align}
	\psi(\phi) = \frac{1}{4}\left( \phi^2 - 1 \right)^2 \qquad \text{with} \qquad
	\psi'(\phi) = \phi^3 - \phi.
	\label{eqn:free_energy}
\end{align}
We choose $u_{r}$ and $L_r$ to be the reference velocity and length, $m_{r}$ is the reference mobility, $\sigma$ is the scaling interfacial tension, $\nu_{r} = \eta_{+}/\rho_{+}$, $\varepsilon$ is the interface thickness, $g$ is gravitational acceleration. 
Then the non-dimensional parameters are as follows: Peclet, $Pe = \frac{u_{r} L_{r}^2}{m_{r}\sigma}$; Reynolds, $Re = \frac{u_{r} L_{r}}{\nu_{r}}$; Weber, $We = \frac{\rho_{r}u_{r}^2 L_{r}}{\sigma}$; Cahn, $Cn = \frac{\varepsilon}{L_{r}}$; and Froude, $Fr = \frac{u_{r}^2}{gL_{r}}$, with $u_{r}$ and $L_r$ denoting the reference velocity and length, respectively. 
Note that the $Cn$ number is the non-dimensional interface thickness in the method. 
We use scaling relation suggested by~\citep{ Magaletti2013} of $1/Pe = 3 Cn^2$ to calculated appropriate $Pe$ for a chosen $Cn$\footnote{See Remark 1 in~\citet{Khanwale2021}.}.   
%
%
%

The system of equations \cref{eqn:nav_stokes} -- \cref{eqn:phi_eqn} has a dissipative law given by: 
\begin{equation}
\d{E_{\text{tot}}}{t} = -\frac{1}{Re}  \int_{\Omega} \frac{\eta(\phi)}{2} \norm{\nabla \vec{v}}_F^2 \mathrm{d}\vec{x} - \frac{Cn}{We} \int_{\Omega}m(\phi) \norm{\nabla \mu}^2  \mathrm{d}\vec{x},
\end{equation}
where the total energy is 
\begin{equation}
E_{\text{tot}}(\vec{v},\phi, t) = \int_{\Omega}\frac{1}{2}\rho(\phi) \norm{\vec{v}}^2 \mathrm{d}\vec{x} + \frac{1}{CnWe}\int_{\Omega} \left(\psi(\phi) + \frac{Cn^2}{2} \norm{\nabla\phi}^2 + \frac{1}{Fr} \rho(\phi) y \right) \mathrm{d}\vec{x}.
\label{eqn:energy_functional}
\end{equation}
The norms used in the above expression are the Euclidean vector norm and the Frobenius matrix norm:
\begin{equation}
\norm{\vec{v}}^2 := \sum_i \abs{v_i}^2 \qquad \text{and} \qquad 
\norm{\nabla\vec{v}}^2_F := \sum_i \sum_j \abs{\frac{\partial v_i}{\partial x_j}}^2.
\end{equation}
This energy functional does not consider energy fluxes into/out of the boundaries. This scenario is representative of closed systems, systems with neutral contact line dynamics, or boundary conditions that do not contribute to energy. The dissipation law represents the underlying thermodynamic consistency of the \cref{eqn:nav_stokes} -- \cref{eqn:phi_eqn} (CH-NS) system. 

For the jet simulation we choose a degenerate mobility $m\left(\phi\right) = 1 - \phi^2 \geq 0$. The choice of degenerate mobility allows for better-bound preservation of the phase-field $\phi$ and minimizes coarsening from Cahn-Hilliard. 

\section{Numerical Method}
We use a semi-implicit, projection-based finite element framework for solving the CHNS system.
We use a block iterative hybrid method. 
We use a projection-based semi-implicit time discretization for the Navier-Stokes equation and a fully-implicit time discretization for the Cahn-Hilliard equation. 
We use a conforming continuous Galerkin (cG) finite element method in space equipped with a residual-based Variational Multiscale (RBVMS) formulation.  The residual-based VMS formulation stabilizes pressure and allows LES-type modeling for coarse mesh resolutions. 
Pressure is decoupled using a projection step, which results in two linear positive semi-definite systems for velocity and pressure.  
We deploy this approach on a massively parallel numerical implementation using parallel octree-based adaptive meshes. 
The overall approach allows using relatively large time steps with much faster time-to-solve than similar fully-implicit methods. The details of the numerical method and its testing and validation for canonical cases are given in~\citep{Khanwale2021}.

\section{Details of the drop in swirling flow case}

The case presented in Figure 4 is a popular test case in the two-phase flow literature proposed by~\citet{Bell1989}. A non-dimensional domain of $1 \times 1$ is selected with a drop of the radius of $0.15$ located initially at (0.5, 75). This circle in this test is deformed with a solenoidal field with a stream function $\psi = (1/\pi)  \sin^2(\pi x)\sin^2(\pi y)$. The advection by the velocity field causes the drop to evolve into a thin filament that spirals toward the center of the vortex (0.5, 0.5). The evolution of thin filaments is very useful for testing the methods for the numerical breakup. 

For the case of presented in Fig 1(a) we use a constant mobility $m(\phi) = 1.0$. A non-dimensional thickness of $Cn = 0.0025$. We use a base mesh refinement of level 5 ($1/2^5$) with a finer refinement near the interface of level 9 ($1/2^9$). For the case in Fig1(b), we use a dynamically reduced non-dimensional interface thickness where the $Cn$ number is reduced from $0.0025$ to $0.001$ with an associated refinement of level 9 ($1/2^9$) and ($1/2^{12}$) respectively. We use a degenerate mobility $m\left(\phi\right) = 1 - \phi^2 \geq 0$ for the dynamic $Cn$ case. ~\cref{fig:swirling_flow_time} shows the time evolution of the two cases (Fig. 1 in the manuscript compares a one-time snapshot for this case). Notice the local decrease in the interface thickness. This local decrease  prevents filament breakup, as seen in~\cref{fig:swirling_flow_time}(b). In contrast, the case with a constant interface thickness in~\cref{fig:swirling_flow_time}(a) undergoes numerical breakup and coarsening from Cahn-Hilliard. 

\input{Tikz/SwirlingFlow.tex}

\section{Detection of drops}
\begin{figure*}
	\centering
	\includegraphics[width=\linewidth]{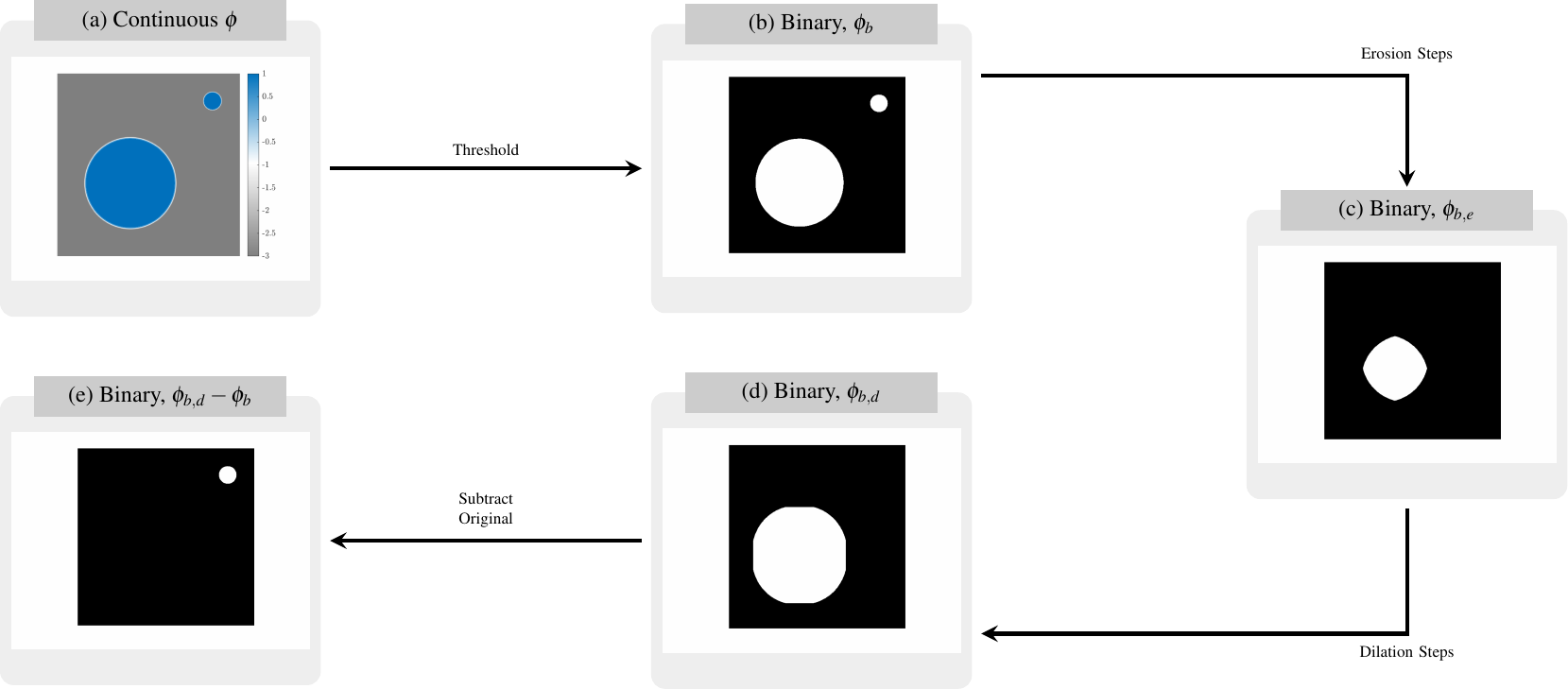}
	\caption{Flow chart describing the algorithm for detecting small droplets. (a) Start with the continuous field $\phi$; (b) Threshold to produce a binary image; (c) Binary after steps of erosion on $\phi_{b}$; (d) Binary after steps of dilation on $\phi_{b,d}$;(e) Binary after original $\phi_{b}$ is subtracted from $\phi_{b,d}$ which shows the identified small drop.}
	\label{fig:schematic}
	\vspace{-0.2in}
\end{figure*}

In the manuscript in Figure 2, we demonstrate the computational technique of detecting a thin filament with a 2D example. Another type of small feature that needs to be detected in atomization simulation is small droplets so that the interface thickness of these drops can be locally decreased. In~\cref{fig:schematic}, we demonstrate the algorithm working to detect a small droplet with a 2D example. 

\section{Time evolution of primary jet atomization}

\begin{figure*}[ht]
		\begin{center}
			\begin{tabular}{|c|c|}
				\hline
				\rotatebox[origin=c]{90}{$t =  \SI{1.648}{\micro\second}$ }
				& 
				\parbox[c]{0.8\linewidth}{
					\includegraphics[
					scale=0.2,trim=0 250 0 250,clip]{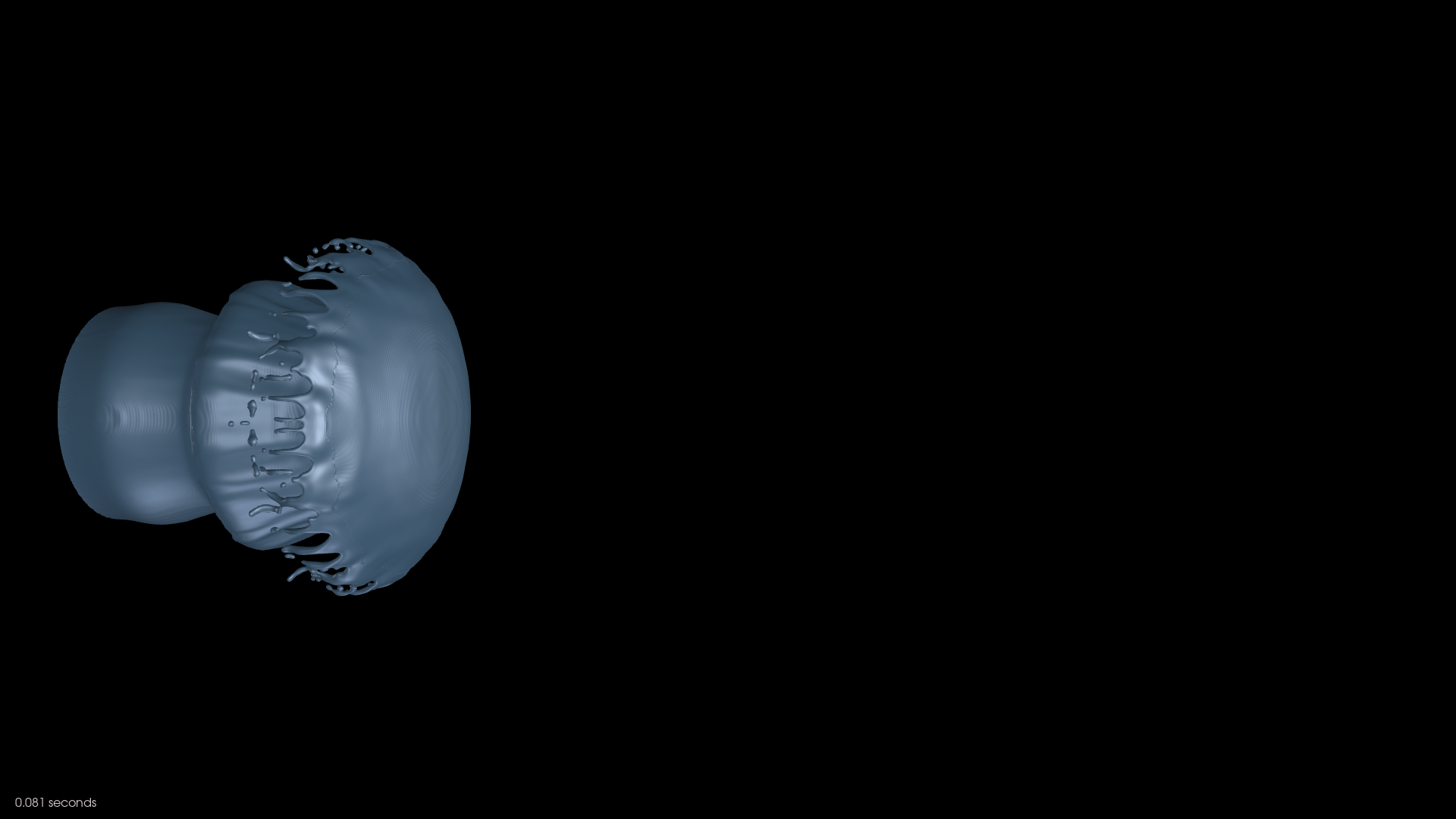}}\\
				\hline
				\hline
				\rotatebox[origin=c]{90}{$t =  \SI{2.824}{\micro\second}$}
				&  
				\parbox[c]{0.8\linewidth}{
					\includegraphics[scale=0.2,trim=0 200 0 200,clip]{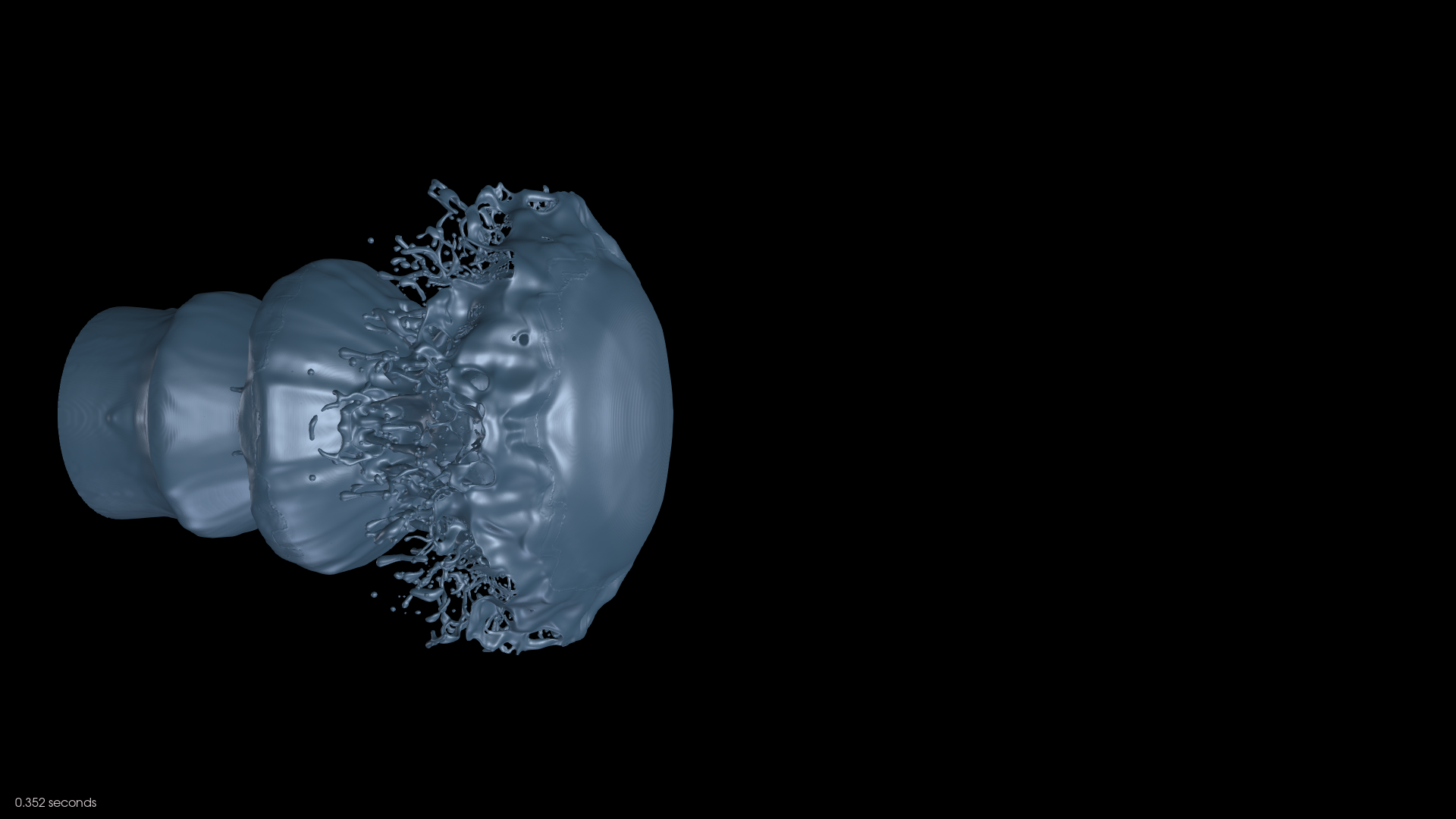}}\\
				\hline
				\hline
				\rotatebox[origin=c]{90}{$t =  \SI{5.384}{\micro\second}$ }
				&  
				\parbox[c]{0.8\linewidth}{
					\includegraphics[scale=0.2,trim=0 100 0 100,clip]{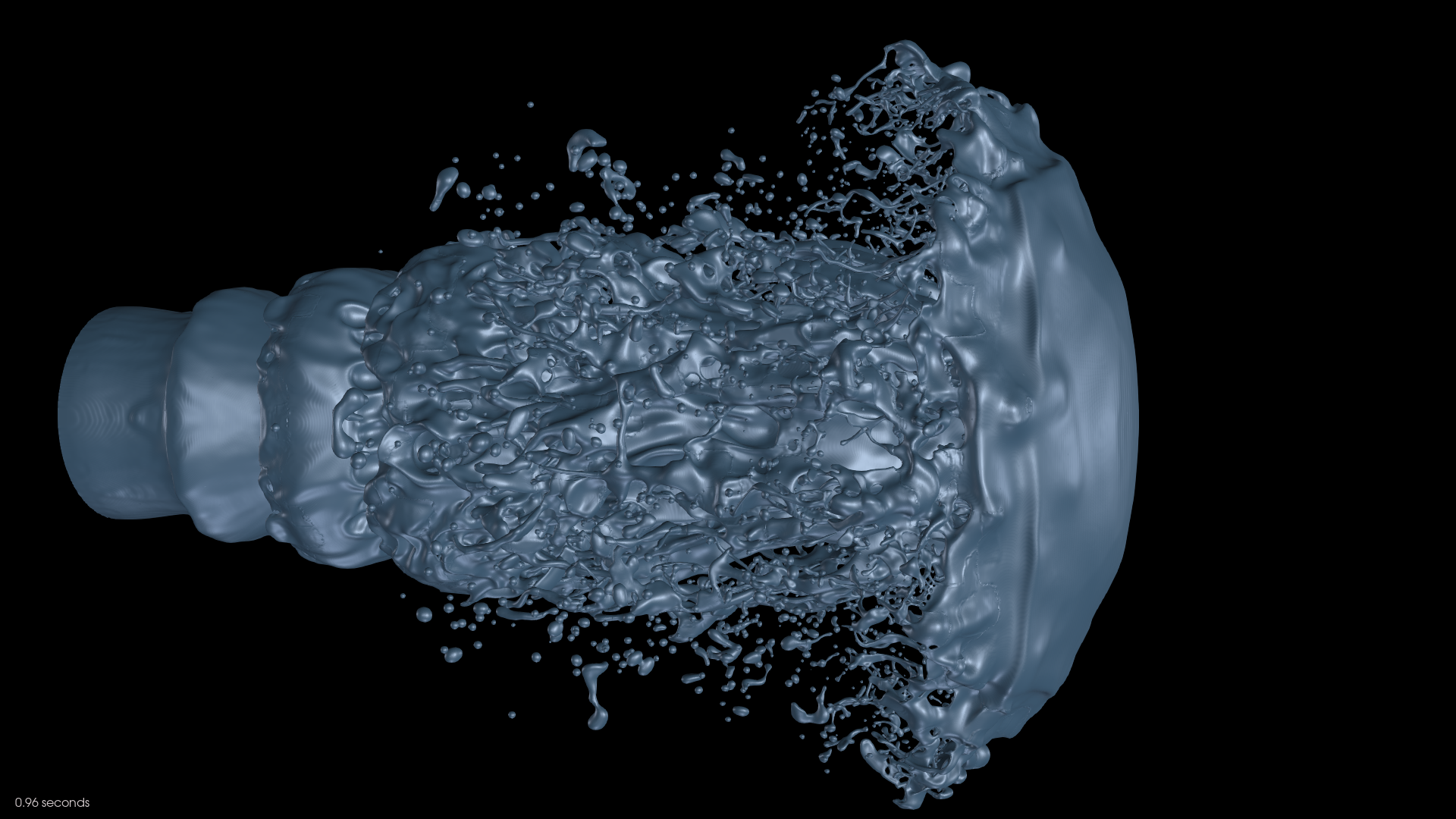}}\\
				\hline
				\hline
				\rotatebox[origin=c]{90}{$t =  \SI{6.216}{\micro\second}$}
				&  
				\parbox[c]{0.8\linewidth}{
					\includegraphics[scale=0.2,trim=0 0 0 0,clip]{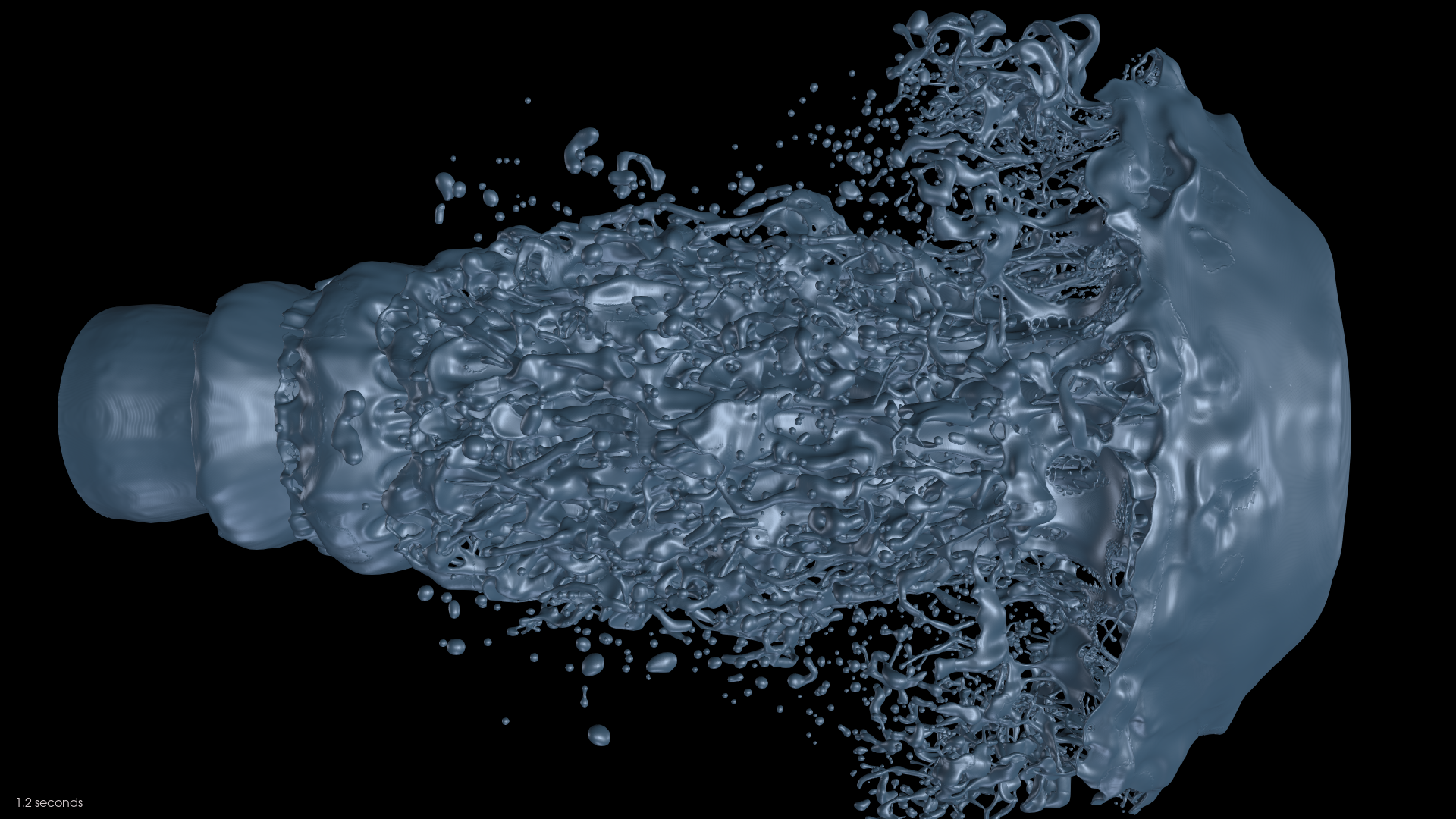}}\\
				\hline
			\end{tabular}		
		\end{center}
		\caption{Time evolution of the primary jet atomization case. Note that 1 non-dimensional time unit = $$\SI{1.0}{\micro\second}$$ }
		\label{fig:jet_zoom_out}
\end{figure*}

\begin{figure}[ht]
	\centering	
	\begin{tabular}{p{0.32\textwidth}p{0.32\textwidth}p{0.32\textwidth}}
		\subfigure [$t = \SI{5.9605}{\micro\second}$] {
			\includegraphics[width=\linewidth]{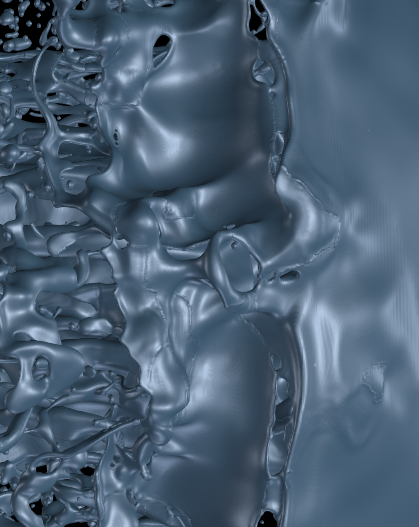}
			\label{subfig:rt3d_mesh_At15_snap_1}
		} &
		\subfigure [$t = \SI{5.9925}{\micro\second}$] {
			\includegraphics[width=\linewidth]{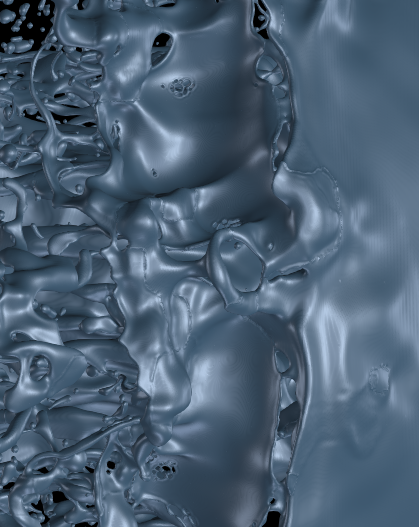}
			\label{subfig:rt3d_mesh_At15_snap_2}
		} & 
		\subfigure [$t = \SI{6.0245}{\micro\second}$] {
			\includegraphics[width=\linewidth]{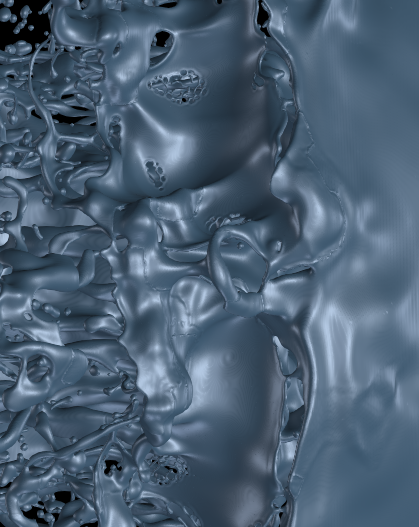}
			\label{subfig:rt3d_mesh_At15_snap_3}
		} \\
		
		\subfigure [$t = \SI{6.0565}{\micro\second}$] {
			\includegraphics[width=\linewidth]{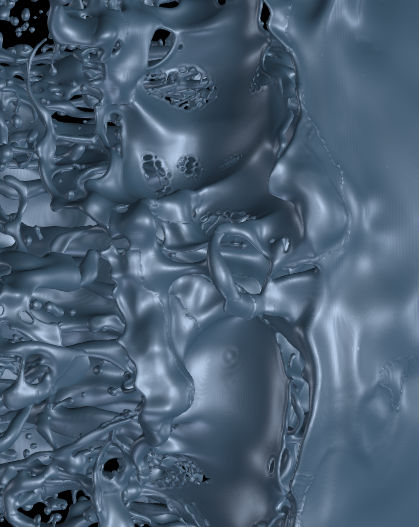}
			\label{subfig:rt3d_mesh_At15_snap_4}
		} &
		\subfigure [$t = \SI{6.0885}{\micro\second}$] {
			\includegraphics[width=\linewidth]{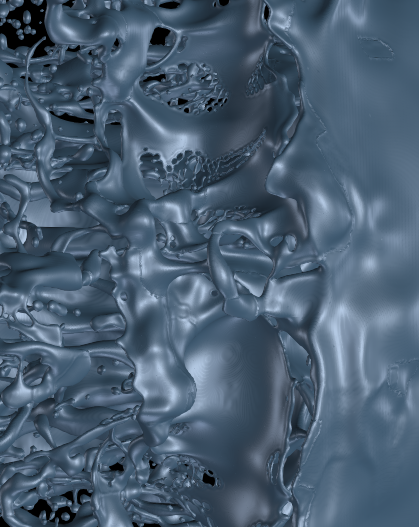}
			\label{subfig:rt3d_mesh_At15_snap_5}
		} & 
		\subfigure [$t = \SI{6.1205}{\micro\second}$] {
			\includegraphics[width=\linewidth]{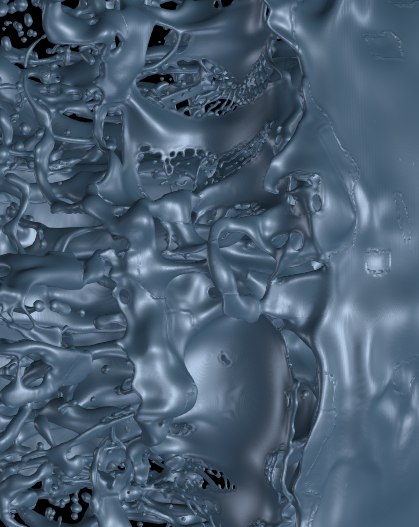}
			\label{subfig:rt3d_mesh_At15_snap_6}
		} \\
		
		\subfigure [$t = \SI{6.1525}{\micro\second}$] {
			\includegraphics[width=\linewidth]{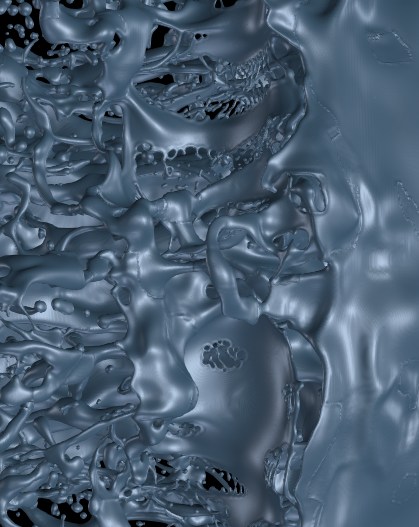}
			\label{subfig:rt3d_mesh_At15_snap_7}
		} &
	\end{tabular}
	\caption{Time evolution of the primary jet atomization case: Zoom in to rupture of sheets.}
	\label{fig:jet_zoom_in_sheets}
\end{figure}

In~\cref{fig:jet_zoom_out}, we show snapshots of the jet atomization with the full jet in the view. This is to show the evolution of the jet from Figure 4 of the manuscript. \Cref{fig:jet_zoom_in_sheets} shows the time evolution of sheet ruptures zoomed in the area shown in Figure 4 of the manuscript. Note that in~\cref{fig:jet_zoom_in_sheets}, we see multiple sheet rupture events in a short period from 5.9605 to 6.1525 $\SI{}{\micro\second}$. 

\bibliography{references}

%% file: preamble.tex
\usepackage{graphicx}
\usepackage{dcolumn}
\usepackage{xcolor}
\usepackage{stix}
\usepackage{bm}

\usepackage{natbib} 
\usepackage{float} 
\usepackage{subfigure}
\usepackage{siunitx}

\usepackage{hyperref}

\hypersetup{
	colorlinks,
	linkcolor={blue!50!black},
	citecolor={blue!50!black},
	urlcolor={blue!80!black},
	anchorcolor = {blue!80!black},
	filecolor = {blue!80!black},
	menucolor = {blue!80!black},
	runcolor = {blue!80!black}
}
\usepackage[capitalise]{cleveref}

\usepackage{pgfplots}
\pgfplotsset{compat=newest}
\usetikzlibrary{arrows.meta}
\usepgfplotslibrary{patchplots}
\usepackage{grffile}
\usepackage{amsmath}

\definecolor{mycolor1}{rgb}{0.00000,0.44700,0.74100}%
\definecolor{mycolor2}{rgb}{0.63500,0.07800,0.18400}%
\usepackage{xspace}
\let\storeBeta=\beta
\renewcommand\beta{\relax\ifmmode{\storeBeta}\else{$\storeBeta$}\fi\xspace}

\let\storeAlpha=\alpha
\renewcommand\alpha{\relax\ifmmode{\storeAlpha}\else{$\storeAlpha$}\fi\xspace}

\let\baraccent=\= 
\renewcommand{\=}[1]{\stackrel{#1}{=}} 

\newcount\colveccount
\newcommand*\colvec[1]{
	\global\colveccount#1
	\begin{pmatrix}
		\colvecnext
	}
	\def\colvecnext#1{
		#1
		\global\advance\colveccount-1
		\ifnum\colveccount>0
		\\
		\expandafter\colvecnext
		\else
	\end{pmatrix}
	\fi
}

%% file: Tikz/SwirlingFlow.tex
\newcommand \figwidthCase{10}
\begin{figure*}[ht]
 \begin{center}
  \begin{tabular}{|c|c|c|c|c|c|}
  \hline
  & t = 0
  & t = 2
  & t = 4
  & t = 6
  & t = 8
  \\
  \hline
\rotatebox[origin=c]{90}{Constant $\epsilon$ (Fig 1(a))}
    & 
 \parbox[c]{\figwidthCase em}{
      \includegraphics[scale=0.15
     ,trim=6.5in 0 4.0in 0,clip
     ]{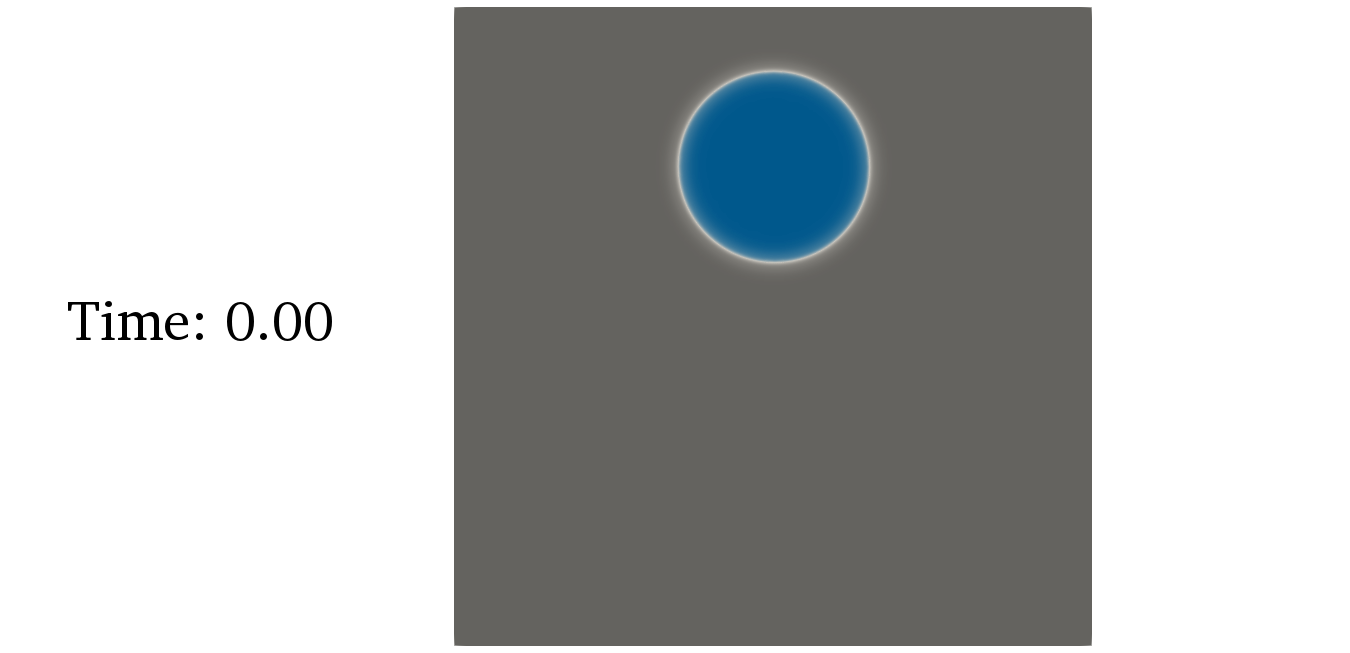}}
    & 
 \parbox[c]{\figwidthCase em}{
      \includegraphics[
      scale=0.15,trim=6.5in 0 4.0in 0,clip
      ]{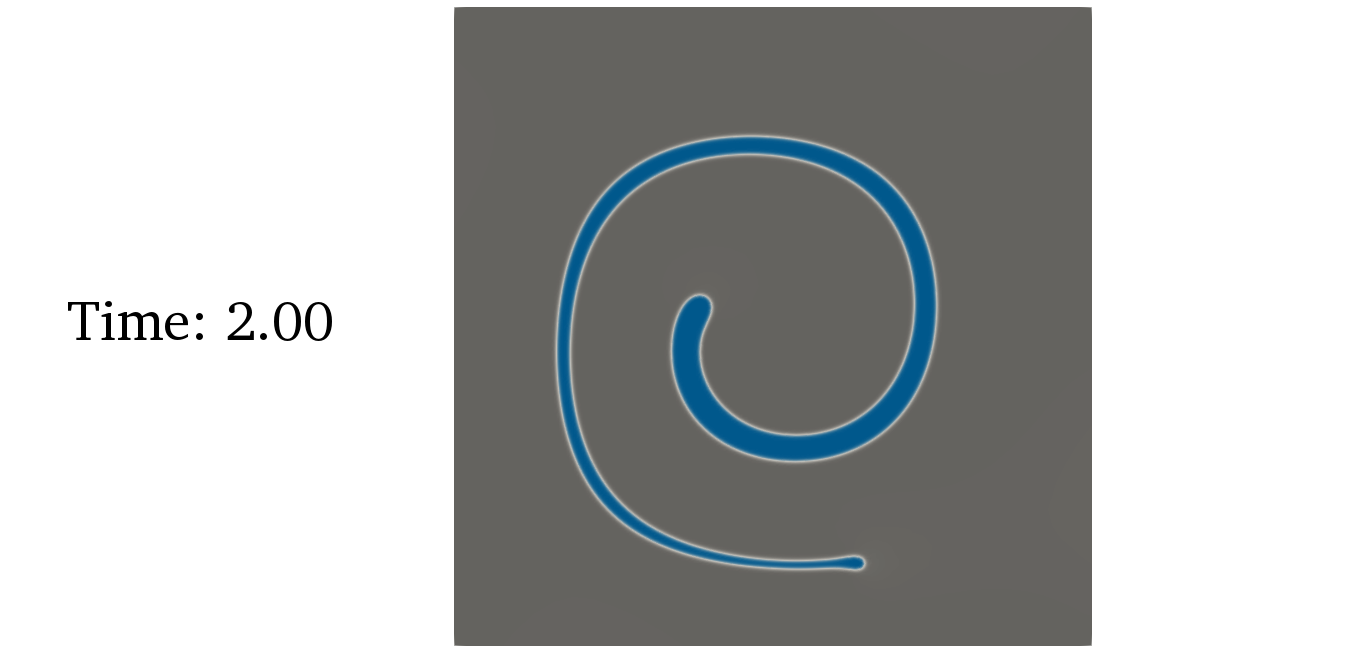}}
      &
      \parbox[c]{\figwidthCase em}{
      \includegraphics[scale=0.15,trim=6.5in 0 4.0in 0,clip
      ]{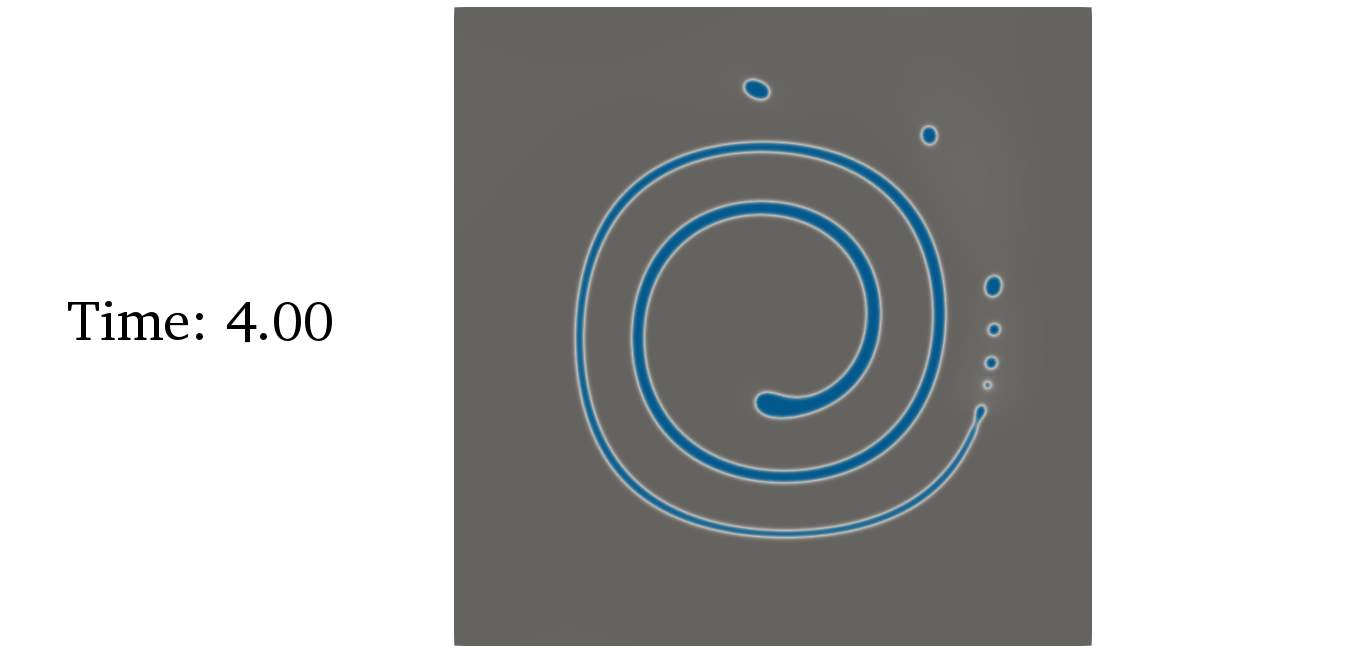}}
      &
      \parbox[c]{\figwidthCase em}{
      \includegraphics[
  scale=0.15,trim=6.5in 0 4.0in 0,clip
      ]{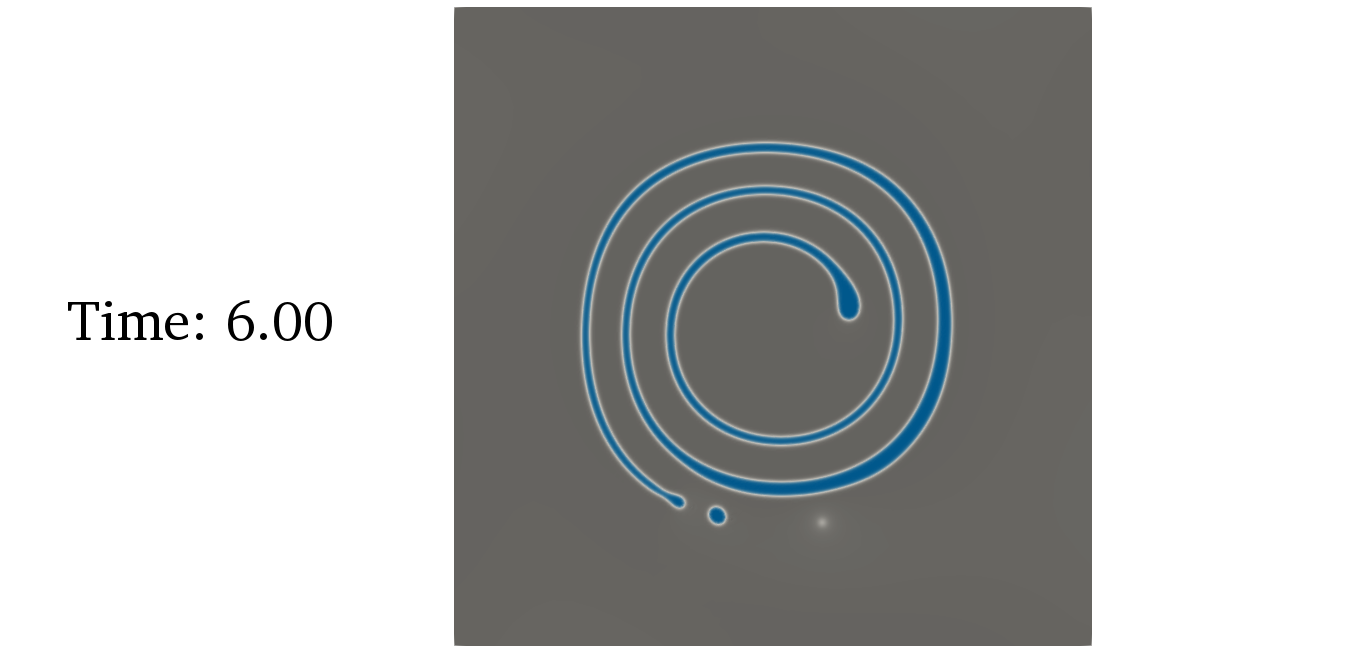}}
      &
      \parbox[c]{\figwidthCase em}{
      \includegraphics[scale=0.15,trim=6.5in 0 4.0in 0,clip
      ]{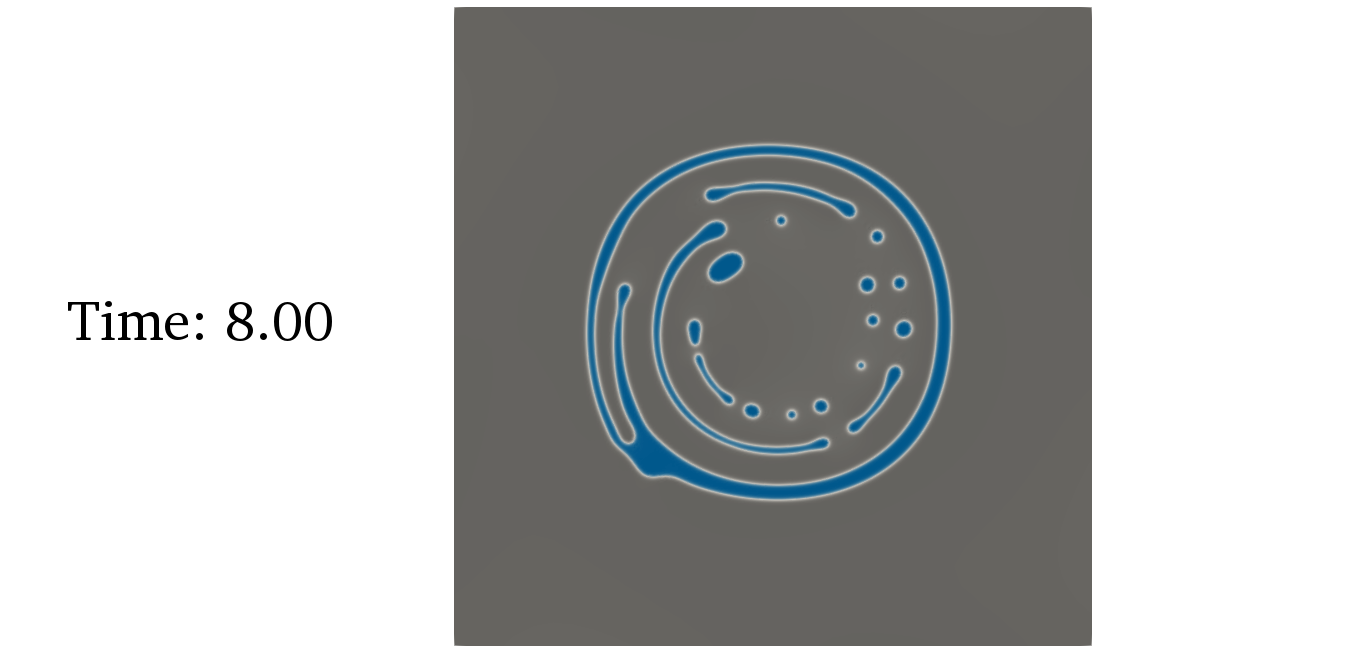}}
      \\
      \hline
      \rotatebox[origin=c]{90}{Dyanmic $\epsilon$ (Fig 1(b))}
    & 
 \parbox[c]{\figwidthCase em}{
      \includegraphics[scale=0.15,trim=6.5in 0 4.0in 0,clip
     ]{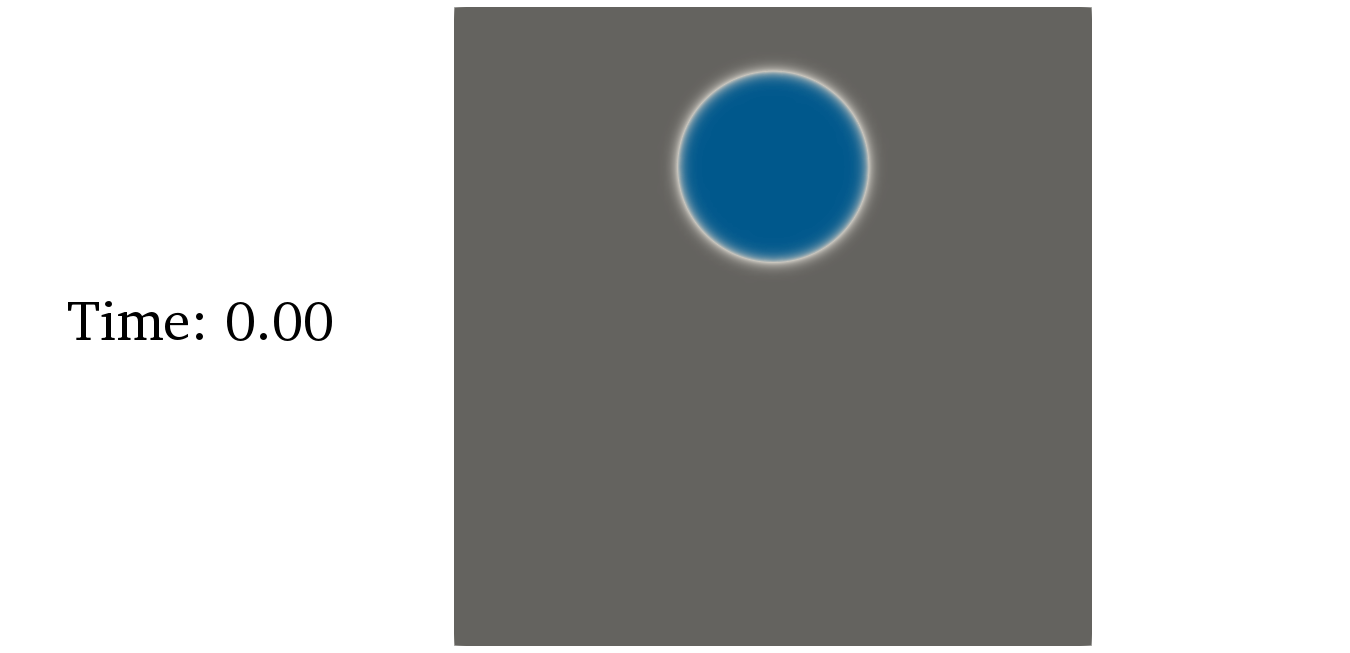}}
    & 
 \parbox[c]{\figwidthCase em}{
      \includegraphics[scale=0.15,trim=6.5in 0 4.0in 0,clip
     ]{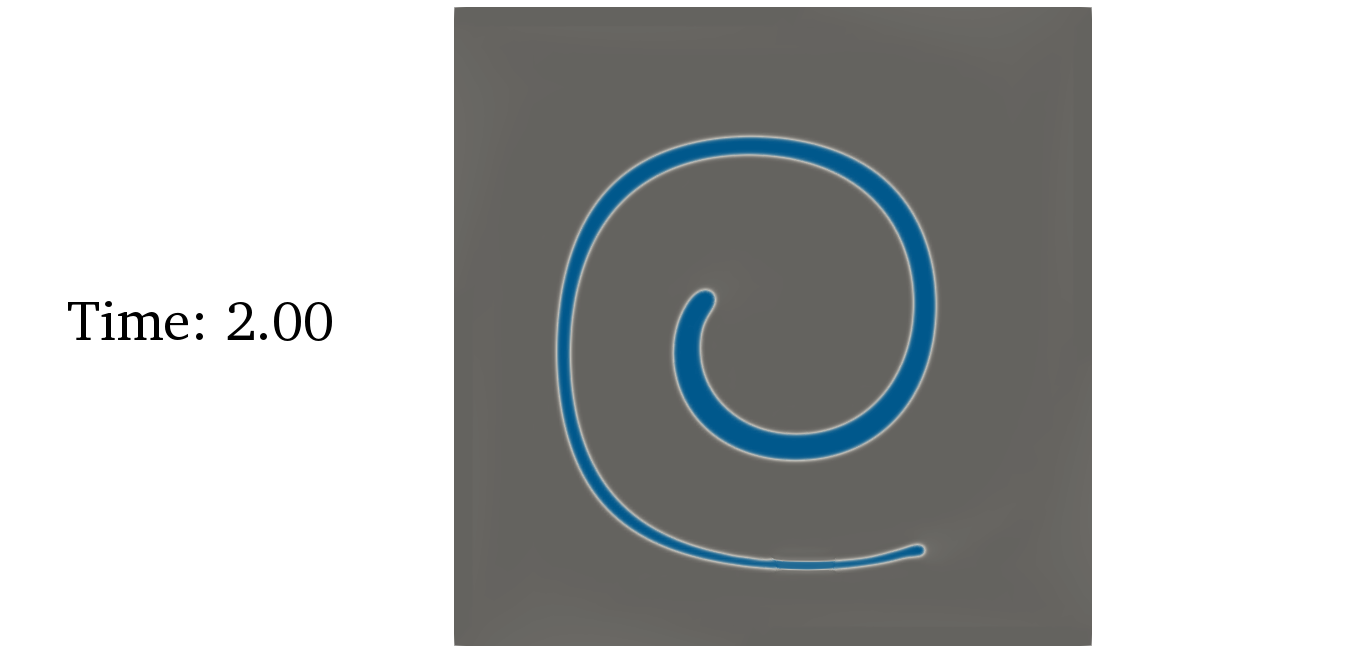}}
      &
      \parbox[c]{\figwidthCase em}{
      \includegraphics[scale=0.15,trim=6.5in 0 4.0in 0,clip
     ]{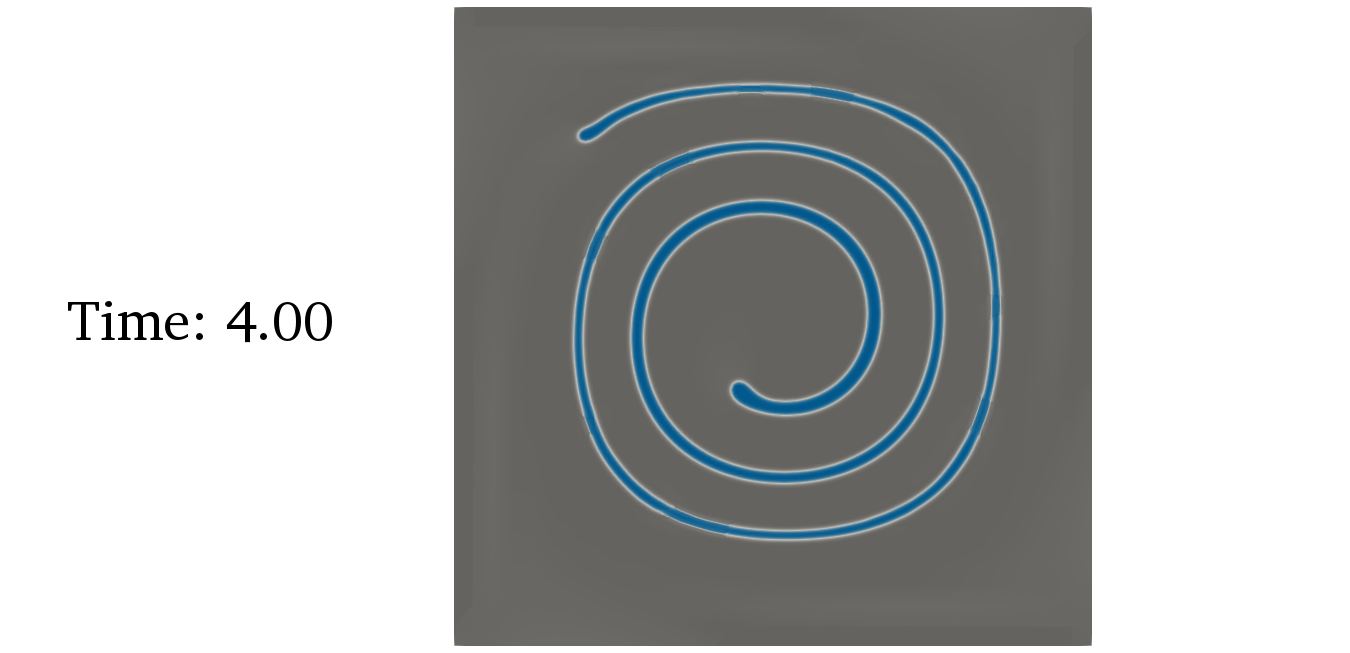}}
      &
      \parbox[c]{\figwidthCase em}{
      \includegraphics[scale=0.15,trim=6.5in 0 4.0in 0,clip
     ]{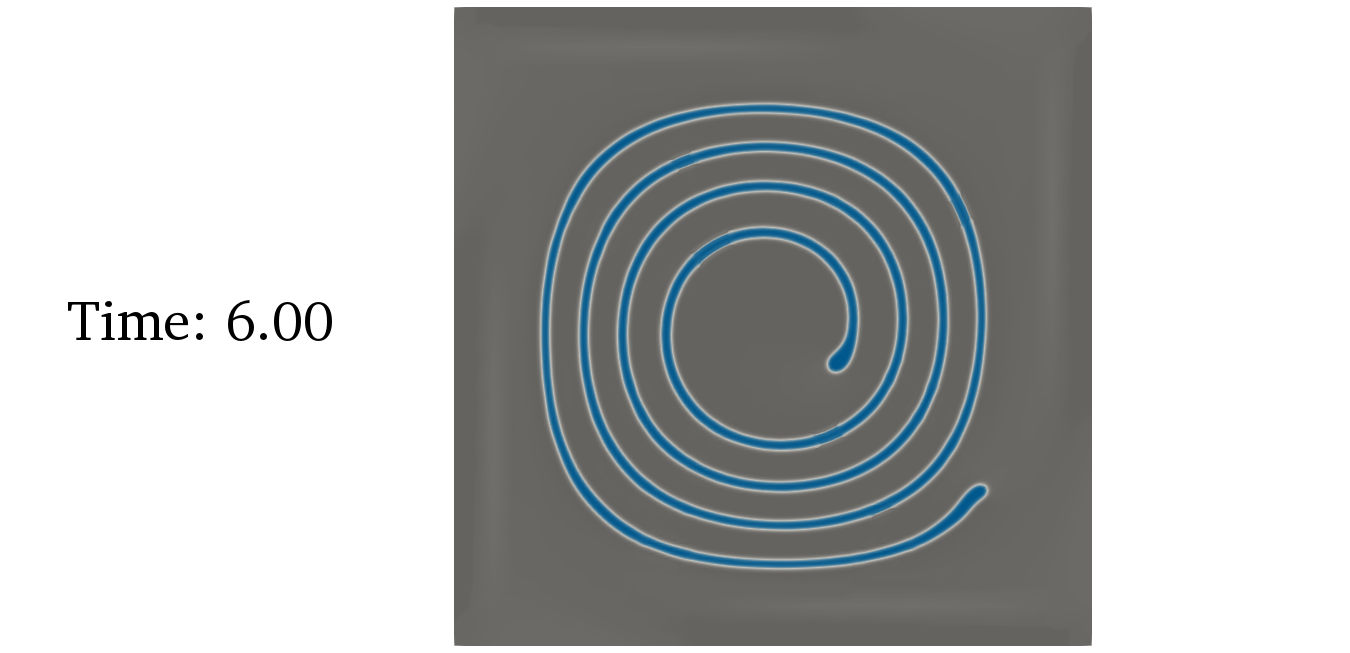}}
      &
      \parbox[c]{\figwidthCase em}{
      \includegraphics[scale=0.15,trim=6.5in 0 4.0in 0,clip
     ]{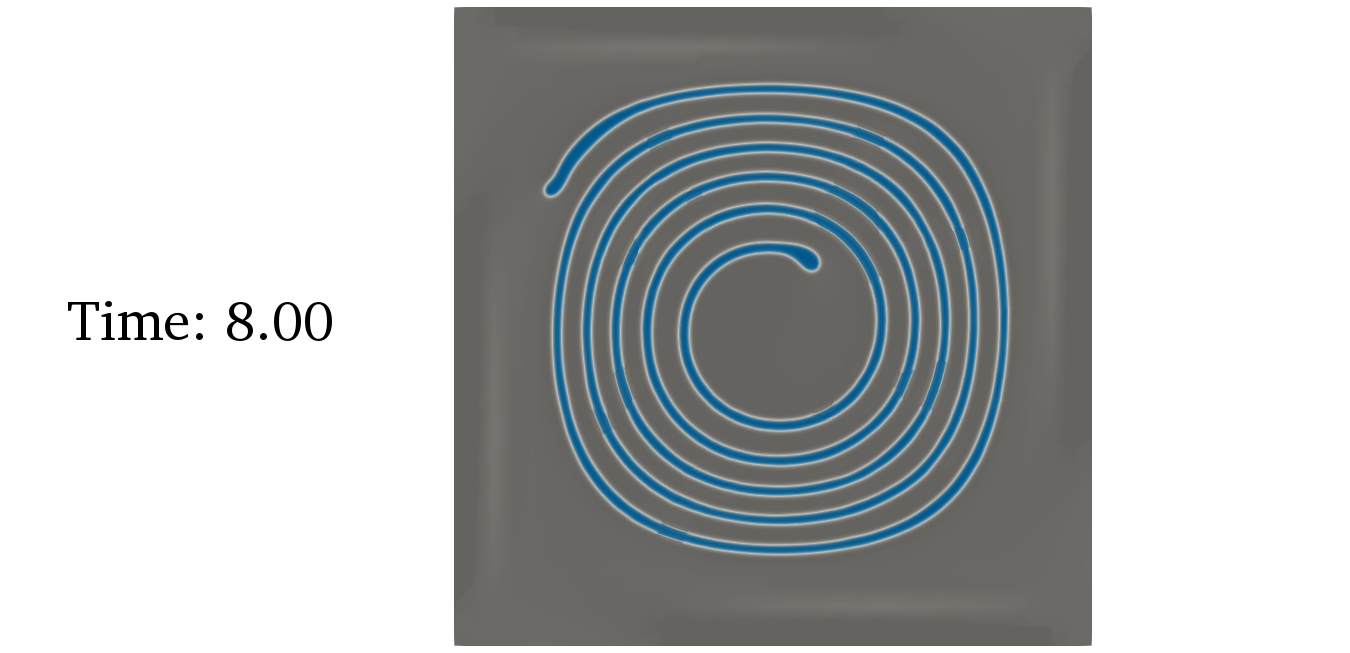}}
      \\
      \hline
  \end{tabular}
  
  \end{center}
  \caption{Time evolution of the cases in Fig 1 in the manuscript.}
  \label{fig:swirling_flow_time}
\end{figure*}